\documentclass[11pt,letterpaper]{article}
\usepackage[margin=1in]{geometry}

\usepackage[utf8]{inputenc}
\usepackage[T1]{fontenc}
\usepackage{microtype}

\usepackage[margin=1in]{geometry}
\usepackage[english]{babel}
\usepackage{amsthm,amsmath,amsfonts,amssymb,amstext}
\usepackage{dsfont}
\newcommand{\mathbbm}[1]{\mathds{#1}}
\newcommand{\indic}{\mathbbm{1}}

\usepackage{subcaption}
\usepackage{nicefrac}
\usepackage{thmtools}
\usepackage{array}

\makeatletter
\@ifundefined{newcounteralias}{}{%
  \renewcommand\thmt@autorefsetup{%
    \@xa\def\csname\thmt@envname autorefname\@xa
      \endcsname\@xa{\thmt@thmname}%
  }%
}
\makeatother

\usepackage{csquotes}
\usepackage{xcolor}
\usepackage{graphicx}

\usepackage{mathtools}
\usepackage{xspace}
\usepackage{enumitem}
\usepackage[linesnumbered,ruled,vlined]{algorithm2e}
\usepackage{multicol}

\usepackage[colorlinks=true, allcolors=blue]{hyperref}
\usepackage[capitalise,nameinlink]{cleveref}
\crefname{appendix}{Appendix}{Appendices}
\Crefname{appendix}{Appendix}{Appendices}
\crefname{subappendix}{Appendix}{Appendices}
\Crefname{subappendix}{Appendix}{Appendices}
\crefname{subsubappendix}{Appendix}{Appendices}
\Crefname{subsubappendix}{Appendix}{Appendices}

\newtheorem{theorem}{Theorem}

\newtheorem{corollary}[theorem]{Corollary}

\newtheorem{remark}[theorem]{Remark}

\newcommand{\eps}{\varepsilon}

\usepackage{xspace}

\newcommand{\opt}{{\mathrm{OPT}}\xspace}

\newcommand{\EE}{\mathbb{E}}

\renewcommand{\Pr}{\mathbb{P}}

\title{Distributionally Robust Search: The Power of Information\\ and the Limits of Moments} 
\author{Nicole Megow\thanks{Faculty of Mathematics and Computer Science, University of Bremen, Germany. Email: \texttt{nmegow@uni-bremen.de}}
\and Cliff Stein\thanks{Department of Industrial Engineering and Operations Research, Columbia University, New York, United States.  Email: \texttt{cliff@ieor.columbia.edu}}
\and Jos\'e Verschae\thanks{Pontificia Universidad Cat\'olica de Chile, Institute for Mathematical and Computational Engineering, Chile. Email: \texttt{jverschae@uc.cl}}}
\date{} %

\begin{document}
\pagestyle{empty}
\pagenumbering{gobble}

\maketitle
\thispagestyle{empty}

\begin{abstract} 

We study binary search for a target drawn from a distribution on $n$ ordered elements. The algorithm receives only partial information about this distribution, and its cost is the expected number of comparison queries. To quantify the loss from this limited information, we use a distributionally robust framework based on uncertainty sets: the algorithm knows only an uncertainty set containing the true distribution, and must minimize its worst-case approximation ratio with respect to an optimal strategy that knows the distribution. Different collections of uncertainty sets capture qualitatively different forms of information, such as quantiles, moments, general linear measurements, or approximate CDFs. This framework allows us to compare these forms of information on equal terms, and thus quantify both how much and what form of information is needed for near-optimal search.

Our main technical result separates moments from general linear measurements. Although $n-1$ moments determine the distribution exactly, for every constant $\delta\in(0,1)$, the first $n^{1-\delta}$ exact moments can be as uninformative for search as no distributional information at all: the worst-case approximation ratio remains $\Omega(\log n)$. In contrast, for every constant $\eps\in(0,1)$, just $O(n^\eps\log^4 n)$ suitably chosen linear measurements suffice for an $O(1/\eps)$-approximation. The limitation therefore lies in the choice of measurements, rather than linearity itself. Using LP duality and polynomial bounds of Rakhmanov and Bernstein, we construct distributions with identical moments whose mass is concentrated at different positions. For the upper bound, we combine dyadic sketches and Johnson--Lindenstrauss embeddings with a truncated median-search algorithm that converts uniform CDF approximation into search guarantees.

Measuring information in bits, we prove a lower bound of $\Omega(1/\eps)$ on the approximation ratio for summaries of any form using at most $\frac12 n^\eps\log n$ bits, for every sufficiently small constant $\eps>0$. Quantiles match this up to constant factors. For Kolmogorov balls of radius $n^{-\eps}$ the optimal approximation ratio is $\Theta(1/\eps)$.
\end{abstract}

\clearpage
\pagenumbering{arabic}
\pagestyle{plain}
\tableofcontents

\newpage
\section{Introduction}

How much information about a distribution is needed to make good decisions under uncertainty? We study this question by considering a distributionally robust framework based on uncertainty sets, and apply it to the fundamental problem of searching on an ordered array. Specifically, we ask which information about the target distribution allows us to search nearly as efficiently as if the distribution were completely known.

Consider an unknown target $t$ drawn from a distribution $p=(p_0,\ldots,p_{n-1})$ on $[n]_0:=\{0,\ldots,n-1\}$. A query at a vertex $v$ reveals whether $t<v$, $t=v$, or $t>v$. We seek an adaptive search strategy minimizing the expected number of queries. Without information about $p$, binary search finds every target in $O(\log n)$ queries. When $p$ is known, an optimal strategy can be computed by dynamic programming~\cite{knuthOptimumBinarySearch1971}. 
Moreover, distribution-guided bisection achieves expected cost $O(1+H(p))$, within a constant factor of the optimum $\opt(p)$, where $H(p)$ %
is the entropy of~$p$~\cite{mehlhornNearlyOptimalBinary1975}.
Distributional knowledge can therefore reduce the expected search cost from logarithmic to constant.

We study how much of this advantage can be retained with limited knowledge of $p$. In our distributionally robust model, the available information specifies an uncertainty set $U$ containing $p$, and the search strategy must perform well for every distribution in $U$. For example, if we observe a sketch of $p$, such as its first $k$ moments, then $U$ is the set of all distributions with the same sketch. %
We compare the strategy's expected cost under each $p\in U$ with $\opt(p)$: the benchmark knows the true distribution, whereas our strategy knows only $U$; neither knows the realized target. A collection $\mathcal{U}$ of uncertainty sets describes a form of partial information. Our central question is:

\begin{quote}
\emph{How much and what kind of distributional information is needed to obtain a constant-factor approximation for search?}
\end{quote}

We first measure information in bits, allowing arbitrary encodings of the distribution. This gives a baseline for the \emph{amount} of information needed, independently of how it is represented. 

We then ask how the \emph{form} of the information affects search. A natural class of summaries consists of linear measurements $\sum_t a_t p_t$, with coefficients chosen independently of $p$. Such measurements can be estimated from samples by averaging the corresponding coefficients. %
We ask how many measurements suffice and how their choice affects the achievable approximation. Here we count exact real-valued measurements, rather than the bits needed to represent them.

Moments are a particularly natural class of linear measurements. The first $k$ moments are $\mu_i(p)=\sum_t p_t t^i$ for $i=1,\ldots,k$. Low-order moments have familiar interpretations: %
they determine the mean, variance, skewness, and kurtosis. 
For empirical distributions, moments can be maintained through sums of powers, which are easily updated under insertions and deletions and combined across datasets. Moments therefore provide interpretable, efficiently maintainable summaries of increasing informativeness; on $[n]_0$, the first $n-1$ moments determine $p$ uniquely.

Moment information is widely used to describe distributional uncertainty in robust optimization~\cite{scarf1958minmax,delageYe2010} and mechanism design~\cite{CarrascoLKMMM18,CheZ21,BachrachT22,WangLZ2024}, and its informativeness for optimal stopping has recently been studied~\cite{CorreaCLVZ26}.

For search, however, choosing a distribution consistent with the observed moments does not by itself justify using its optimal strategy: the strategy must perform well for every distribution with those moments. 
Thus, we ask
how many moments are needed for constant-factor approximate search, and how their power compares with that of the same number of arbitrary linear measurements.

\subsection{Our Contributions and Techniques}

We introduce an uncertainty-collection framework for studying how the amount and form of distributional information affect search. 
We measure a collection $\mathcal{U}$ of uncertainty sets by its approximability factor $\alpha(\mathcal{U})$: the worst-case approximation ratio of the best strategy that knows only the uncertainty set $U$, compared with an optimal strategy that knows the true distribution $p\in U$ (\Cref{sec:uncertainty-model}).
 The framework accommodates finite encodings, quantiles, moments, linear measurements, and approximate cumulative distribution functions (CDFs).

We ask which uncertainty collections achieve $\alpha(\mathcal{U})=O(1/\eps)$,
for some positive constant $\eps$, and how much information is required. Our upper bounds use a common sufficient condition: a nominal distribution whose CDF approximates that of every distribution in the uncertainty set uniformly within $n^{-\eps}$, or equivalently, whose Kolmogorov distance $d_K$ to each such distribution is at most $n^{-\eps}$. We consider a natural algorithm, \textsc{Coarse Greedy}, that runs the classical median-search algorithm analyzed by Mehlhorn~\cite{mehlhornNearlyOptimalBinary1975} on the nominal distribution, truncating it at a depth determined by the available accuracy guarantee before switching to ordinary binary search. This yields an $O(1/\eps)$-approximation (\Cref{thm:CoarseGreedy}). Thus, uniform CDF approximation provides a common route from distributional summaries to search guarantees.

\medskip \noindent \textbf{Information-Theoretic Limits\ }
We first establish a lower bound that applies to every encoding of the distribution into a finite bit string, regardless of how the information is computed or represented. %

\begin{theorem}[Information-theoretic lower bound, informal; see \Cref{thm:informationLowerBound}]\label{thm:IntroInfTheoreticalLowerBound}
Let $\eps>0$ be a small constant.   For  $n$  large enough and for  every sketch $\sigma$ whose output uses at most $\frac12 n^\eps\log n$ bits, every algorithm that only has access to $\sigma(p)$ has approximation ratio $\Omega(1/\eps)$.
\end{theorem}

This lower bound is tight up to constant factors: storing $n^\eps$ quantiles uses $O(n^\eps\log n)$ bits and determines the CDF within Kolmogorov distance $n^{-\eps}$, giving an $O(1/\eps)$-approximation through \textsc{Coarse Greedy}.

The proof of the lower bound uses distributions that place equal mass on one independently and uniformly chosen position in each of $n^\eps$ consecutive blocks. If the distribution is known, searching its support takes $O(\eps\log n)$ queries. The key step is to show that a short encoding leaves enough conditional entropy in the support positions that every strategy using this encoding incurs expected cost $\Omega(\log n)$ on some compatible distribution.

We extend this lower bound to arbitrary uncertainty collections, including collections with overlapping sets (\Cref{cor:Coveringlowerbound}). The key observation is that a small cover of the simplex by uncertainty sets induces a short encoding: record the index of a covering set containing the distribution. %
As an application, the quantile construction provides a cover by Kolmogorov balls. Together with \textsc{Coarse Greedy}, this yields an approximability factor of $\Theta(1/\eps)$ for balls of radius~$n^{-\eps}$~(\Cref{thm:kolmogorovBallLowerBound}).

\medskip \noindent \textbf{Search Under Moment Information\ }
We now turn to the interpretable summaries defined by moments. The first $k$ moments define the linear sketch $\sigma_k(p):=(\sum_{t\in[n]_0}t^i p_t)_{i=0}^k$, whose uncertainty collection consists of infinitely many lower-dimensional polytopes $\{p:\sigma_k(p)=\bar\mu\}$. We include the constant zeroth moment $\mu_0(p)=1$ for convenience. Since we count moments rather than the bits needed to represent their values, the finite-bit lower bound does not directly apply. Any linear sketch needs $n^{\Omega(\eps)}$ measurements for an $O(1/\eps)$-approximation (\Cref{thm:linearSketchlowerBound}). This makes $n^\eps$ moments a natural benchmark, up to a constant in the exponent. Our main result shows a much stronger limitation: $\Omega(n/\log n)$ moments are needed for a constant approximability factor (\Cref{thm:momentSearchSeparation}).

\begin{theorem}[Moment lower bound, informal]\label{thm:IntroMomentsLower}
For large $n$ and $k\le cn/\log n$, with $c>0$ sufficiently small, the first $k$ moments induce an uncertainty collection with approximability factor $\Omega\bigl(\log\frac{n}{k\log n}\bigr)$.
\end{theorem}

In particular, for every constant $\delta\in(0,1)$, even $k=n^{1-\delta}$ moments leave an approximability factor of $\Omega(\log n)$, which is trivially obtained by ordinary binary search without any distributional information. Thus, a polynomially sublinear number of moments can be as uninformative for search as no information at all. In contrast, $n-1$ moments determine the full distribution exactly and hence permit an optimal search strategy. Yet $O(n^\eps\log^4 n)$ suitably chosen linear measurements already suffice for an $O(1/\eps)$-approximation (\Cref{cor:searchFromDyadicJL}).

The key ingredient in the proof of \Cref{thm:IntroMomentsLower} is \Cref{prop:2Distros}: two distributions $p,p'$ can share the same first $k$ moments but have peaks of mass at least $1-\gamma$ at different positions $m,m_0$, where $\gamma=O(k|m-m_0|/n)$; see \Cref{2Distros}. In particular, their Kolmogorov distance is at least $1-2\gamma$, so matching moments alone need not determine the CDF accurately. Intuitively, a search strategy that only knows the $k$ moments cannot distinguish between $p$ and $p'$, and hence does not know whether to search at $m$ or at $m_0$. 
To construct them, we start with point masses at $m$ and $m_0$ and seek a signed correction that compensates for the difference in their moments. A correction with small $\ell_1$-norm lets us match the moments by adding little mass to either point mass and then normalizing, so both peaks remain large. The correction problem is an LP whose moment constraints have a Vandermonde coefficient matrix. Its dual therefore has a polynomial interpretation: the minimum correction norm equals the largest increment between $m$ and $m_0$ of a degree-$k$ polynomial bounded on the grid. The key is that such a polynomial cannot change too rapidly in the central part of the domain.  Using Rakhmanov's theorem~\cite{rakhmanov2007} to control  its magnitude between grid points, and Bernstein's inequality to bound its derivative, we can bound  the dual objective, yielding the small correction and hence the two distributions.

To obtain the search lower bound, we combine these corrections to construct $R+1$ distributions with the same first $k$ moments, each concentrated at a different position (\Cref{prop:manyDistrosSameMoments}). Choosing $R+1$ consecutive peak positions with $R=\Theta(\sqrt{n/(k\log n)})$ keeps the mass outside each peak at most $1/(4\log n)$. A strategy knowing the distribution queries its peak first and otherwise uses binary search, giving $\opt=O(1)$. A strategy observing only the common moments must accommodate all $R+1$ possible peaks. A counting argument forces expected cost $\Omega(\log(R+1))$ on one of these distributions, yielding the claimed $\Omega(\log(n/(k\log n)))$ approximation lower bound.

Using the same LP-duality framework, we give proofs of approximate and exact CDF recovery from moments. Approximate recovery holds when one distribution consistent with the moments has bounded atoms, and exact recovery holds when one such distribution has small support.

\begin{theorem}[Recovery from moments, informal]
\label{thm:IntroMomentsUpper}
Let $p$ and $\pi$ be distributions on $[n]_0$ with the same first $k$ moments.
\begin{enumerate}\addtolength{\itemsep}{-1ex}
\item[(a)] If $\pi_t\le L$ for every $t$, then the Kolmogorov distance is $d_K(p,\pi)=O(L\frac{n}{k}\log n)$.
\item[(b)] If $\pi$ has at most $\ell$ support points and $k\ge2\ell$, then $p=\pi$.
\end{enumerate}
\end{theorem}

%
The exact recovery statement in part~(b) is known~\cite[Lemma~4(2)]{wuYang2020}.
LP duality expresses the largest CDF value consistent with the moments as the minimum expectation of a polynomial of degree at most $k$ majorizing the corresponding threshold indicator on the grid. For part~(a), we use Chebyshev polynomials to approximate this indicator from above with transition width $O((n/k)\log n)$ and negligible additional loss (\Cref{lem:cert,thm:momentRecoveryBoundedAtoms}). For part~(b), interpolation gives a dual certificate that agrees with the indicator on the support of $\pi$, making the bound exact (\Cref{prop:momentRecoverySmallSupport}).

Exact recovery in part~(b) yields optimal search. When the bound in part~(a) guarantees CDF error $n^{-\eps}$ with $k\le n$, every compatible distribution has point probabilities $O(n^{-\eps})$, so ordinary binary search already achieves an $O(1/\eps)$-approximation (\Cref{subsec:moment-positive}). If $L=O(n^{-a})$ for constant $a\in(0,1]$, then $k=O(n^{1-a+\eps}\log n)$ moments suffice for CDF error $n^{-\eps}$ for fixed $\eps\in(0,a)$. The separated-peak construction explains why the bound on point probabilities in \Cref{thm:IntroMomentsUpper}(a) matters: without it, distributions with identical moments can have Kolmogorov distance close to one.

\medskip \noindent \textbf{Effective Linear Sketches\ }

It is natural to ask whether the limitations of moments extend to all linear measurements. We show that they do not (\Cref{thm:linearSketchlowerBound,cor:searchFromDyadicJL}).\leavevmode
\begin{theorem}[Linear sketches, informal] \leavevmode
\begin{enumerate}\addtolength{\itemsep}{-1ex}
\item[(a)] Every linear sketch with $k\ge2$ measurements induces an uncertainty collection with approximability factor $\Omega(\log n/\log k)$.
\item[(b)] For every $\eps\in(0,1)$, there exists a linear sketch with $O(n^{\eps}\log^4 n)$ measurements whose
uncertainty collection has approximability factor $O(1/\eps)$.
\end{enumerate}
\end{theorem}
Part (a) implies that $n^{\Omega(\eps)}$ measurements are necessary for an $O(1/\eps)$-approximation, while part (b) shows that this is tight up to a constant in the exponent and polylogarithmic factors. The upper bound adapts dyadic quantile sketches~\cite{gilbert2002summarize,cormode2005improved}, using near-orthogonal vectors from the Johnson--Lindenstrauss lemma~\cite{johnson1984extensions} to sketch interval masses at each level. Every prefix decomposes into $O(\log n)$ dyadic intervals, giving a uniform CDF estimate with error $n^{-\eps/2}$; \textsc{Coarse Greedy} then yields the search guarantee. Thus, suitably chosen linear measurements succeed where even $n^{1-\delta}$ moments fail: the limitation lies in the Vandermonde structure of the moment matrix, rather than linearity itself.

\subsection{Related Literature}

Lin, Luo, and Woodruff~\cite{linLearningAugmentedBinary2022} study learning-augmented binary search trees, built from predicted access frequencies. Dinitz et al.~\cite{dinitz_binary_2024} study binary search with a predicted target distribution. 
Their algorithm achieves expected query complexity $O(H(p)+\log\eta)$, where $H(p)$ is the true distribution's entropy and $\eta$ is its earth mover's distance from the prediction.
Their algorithm interleaves median phases with endpoint searches, without knowing the prediction error beforehand. Our \textsc{Coarse Greedy} uses a known CDF-error bound to fix the truncation depth of median~search. %

Our work connects to the classical moment problem~\cite{schmudgen_moment_2017,karlin1966tchebycheff,kreinNudelman1977}: how do prescribed moments constrain a measure and event probabilities? See Landau~\cite{Landau1987} for its broader mathematical implications. We study this question for distributions supported on a fixed finite grid.

The duality between moments and polynomials goes back to Riesz~\cite{Riesz1923} and Haviland~\cite{Haviland1935}. Together with positivity results such as Putinar's \emph{Positivstellensatz}~\cite{Putinar1993}, it underlies moment-based approaches to polynomial optimization~\cite{Lasserre2001}. Bertsimas and Popescu~\cite{bertsimasPopescu2005} formulate tight probability bounds under moment constraints as optimization problems with polynomial dual certificates.
On our finite domain, bounds on the CDF arise from finite LPs whose duals connect moment information to polynomial approximation.

Polynomial approximation and moment matching also underpin statistical lower bounds for nonsmooth functionals, entropy, and support size~\cite{caiLow2011,wuYang2016,wuYang2019}. There, matching moments of auxiliary distributions makes observations difficult to distinguish. Here, the moments themselves are observed, making distributions with matching moments indistinguishable to the search algorithm.

Correa et al.~\cite{CorreaCLVZ26} study moment information for prophet inequalities, where the goal is to select a large reward from sequential observations. Even knowing all moments yields only a $\Theta(1/\log m)$ competitive guarantee, where $m$ counts the random variables. The two settings are hard for different reasons. Their moments cannot locate a rare atom of large value, which changes the expected maximum by polynomial factors. Search only sees mass, not values, so our lower bound places $\Omega(1)$ mass at different but close positions that the first $k$ moments cannot separate.

Our bit-complexity formulation relates to advice complexity, which asks how many advice bits an online algorithm needs for a given guarantee~\cite{boyarFKLM2017}. The advice may encode arbitrary information, including input properties or a suitable (possibly optimal) solution. Our lower bound (\Cref{thm:informationLowerBound}) applies to every encoding of the distribution, including encodings of search strategies. Beyond counting bits, our uncertainty-set framework compares specific forms of distributional information.

Compact quantile summaries with uniform additive rank-error guarantees are extensively studied in streaming~\cite{greenwald2001space,karnin2016optimal}; see~\cite{cormodeKLTV2023} for relative-error variants. Their CDF approximations for the empirical distribution yield search guarantees through \textsc{Coarse Greedy}. Our bit lower bound concerns search performance directly and applies to arbitrary encodings, even with full access to the distribution and no requirement to approximate its CDF.

\medskip \noindent \textbf{Organization\ }
\Cref{sec:model-coarse-greedy} sets up the search model, the uncertainty model
and the \textsc{Coarse Greedy} reduction. \Cref{sec:information-limits} proves
the bit-complexity and covering-number bounds and the tight characterization of
Kolmogorov balls and quantiles. \Cref{sec:LB} contains the moment results, negative and
positive, while \Cref{sec:linear-sketches} treats general linear sketches.

\section{Model and Search from Approximate CDFs}
\label{sec:model-coarse-greedy}

\subsection{Search Model}
\label{subsec:search-model}

We consider search on the path with vertex set $[n]_0:=\{0,1,\ldots,n-1\}$, with $n\geq 2$.
An unknown target vertex $t\in[n]_0$ is drawn from a distribution
$p=(p_0,\ldots,p_{n-1})$. %
A query at a vertex $v$ reveals whether $t<v$, $t=v$, or $t>v$.
Equivalently, the query either finds the target or identifies the component of the path after removing $v$ that contains it. Logarithms are base~2 unless noted.

A deterministic search strategy is an adaptive procedure that repeatedly
queries vertices until the target is identified.
Equivalently, a strategy is an ordered binary search tree.
Each node corresponds to an interval of possible targets
$I=\{a,\ldots,b\}$ and is labeled by a vertex $v\in I$.
If $a=b$, the target is identified without another query.
Otherwise, the strategy queries $v$, terminates on equality,
and continues on the appropriate nonempty interval
$\{a,\ldots,v-1\}$ or $\{v+1,\ldots,b\}$.

Every strategy must correctly identify every target in $[n]_0$, including targets assigned probability zero by $p$.
For a strategy $S$ and a target $t\in[n]_0$, let $q(S,t)$ denote the number
of queries performed by $S$ before identifying $t$. The expected
cost of $S$ under distribution $p$ is
\[
\operatorname{cost}_p(S):=\mathbb E_{t\sim p}[q(S,t)]
=\sum_{t=0}^{n-1} p_t q(S,t).
\]

We denote the optimal expected search cost by
$\operatorname{OPT}(p):=\min_S \operatorname{cost}_p(S)$, where the minimum is
over all deterministic search strategies on the path.

\subsection{Uncertainty Model}
\label{sec:uncertainty-model}

 Let $\Delta_n = \{p\in \mathbb{R}^n_{\ge0}\,:\, \sum_{i=0}^{n-1}p_i=1\}$ be the simplex of probability distributions on $[n]_0$. A collection of sets $\mathcal{U}\subseteq 2^{\Delta_n}$ that covers $\Delta_n$, i.e., $\bigcup_{U\in \mathcal{U}} U=\Delta_n$, is said to be an \emph{uncertainty collection}. We call a set $U\in \mathcal{U}$ an \emph{uncertainty set}. We are interested in studying which uncertainty collections allow for constant-factor approximations. The approximability factor of an uncertainty collection is defined as
 \begin{equation}\label{eq:alphadef}
  \alpha(\mathcal{U}) := \sup_{U\in\mathcal{U}} \inf_S \sup_{p\in U} \frac{\operatorname{cost}_p(S)}{\opt(p)},
\end{equation}
where $S$ ranges over correct deterministic search strategies.

A natural way of defining an uncertainty collection is via a sketch, which formalizes a summary of the distribution.
Any map $\sigma:\Delta_n\to\mathcal{X}$, for a set $\mathcal{X}$ of
possible messages, is called a sketch. Its uncertainty collection
$\mathcal{U}_{\sigma}$ consists of the nonempty sets
$\{p\in\Delta_n:\sigma(p)=x\}$ for $x\in\mathcal{X}$. 
Thus, the algorithm chooses a search strategy based only on the sketch
$\sigma(p)$ of the unknown target distribution $p$.

Another natural way of defining an uncertainty collection is via balls. Any metric $d(\cdot,\cdot)$ on $\mathbb{R}^n$ together with a parameter $\delta>0$ defines a collection of uncertainty sets $\mathcal{U}_{d,\delta}$, simply by considering all balls of radius $\delta$. That is, each uncertainty set is a ball $B_d(\pi,\delta)=\{p\in \Delta_n \,:\, d(p,\pi)\le \delta \}$ around some nominal distribution $\pi$.  

The cumulative distribution function, or CDF, of $p$ is the vector $P=(P(0),\ldots,P(n-1))$ defined by
$P(t):=\sum_{j=0}^t p_j$ for $t\in [n]_0$. We use the convention
$P(-1):=0$. Thus, for an interval $I=\{a,\ldots,b\}$, its $p$-mass is
$p(I):=\sum_{t=a}^b p_t=P(b)-P(a-1)$. A particularly important distance in our context turns out to be the so-called Kolmogorov distance. Given two distributions $p$ and $\pi$ on $[n]_0$, with CDFs $P$ and $\Pi$,
their Kolmogorov distance is defined as\[
d_K(p,\pi):=\|P-\Pi\|_\infty
=\max_{0\le t\le n-1}|P(t)-\Pi(t)|.
\]
More generally, we will also refer to any nondecreasing vector
$\Pi=(\Pi(0),\ldots,\Pi(n-1))\in[0,1]^n$ with $\Pi(n-1)=1$ as a nominal CDF. We use the
convention $\Pi(-1):=0$. The atom masses induced by $\Pi$ are
$\pi_t:=\Pi(t)-\Pi(t-1)$ for $t\in [n]_0$.

For a nominal distribution $\pi\in\Delta_n$ and radius $\delta>0$,
define the Kolmogorov ball
\[
B_K(\pi,\delta)
:=\{p\in\Delta_n:d_K(p,\pi)\le\delta\}.
\]
The corresponding uncertainty collection is $\mathcal U_{K,\delta}
:=\{B_K(\pi,\delta):\pi\in\Delta_n\}.$

\subsection{Coarse Greedy from Approximate CDFs}
\label{subsec:coarse-greedy}

Our main algorithmic tool is a search strategy computed from a known nominal
distribution $\pi$, which we think of as the center of a Kolmogorov ball
$B_K(\pi,\delta)$ containing the true distribution.
For an interval $I=\{a,\ldots,b\}$ with $\pi(I)>0$, a $\pi$-median is a vertex $m\in I$ satisfying
\[
\pi(\{a,\ldots,m-1\})\le \frac{\pi(I)}2,
\qquad
\pi(\{m+1,\ldots,b\})\le \frac{\pi(I)}2.
\]
Such a vertex always exists. Querying it either finds the target or leaves an interval with at most half the nominal mass.

Classical median search repeatedly queries a probability median of the remaining interval and gives a constant-factor approximation when the distribution is known~\cite{mehlhornNearlyOptimalBinary1975}. We run this strategy using the nominal distribution $\pi$, but truncate it before switching to binary search. The median phase exploits available information; binary search limits the cost when that information~is~too~coarse. %

\medskip \noindent \textbf{Algorithm \textsc{Coarse Greedy}}
Given $\varepsilon\in(0,1]$, the algorithm starts with $I=[n]_0$ and performs at most
$
h:=\lceil\varepsilon\log n\rceil
$
median queries with respect to $\pi$. After each query, it either terminates on equality or continues on the side containing the target. It also terminates whenever the remaining interval is a singleton.
After $h$ median queries, or earlier if $\pi(I)=0$, it finishes
by standard binary search on the remaining interval, independently of the distributional information.
\medskip 

To explain the truncation depth, write the CDF-error bound as $\delta=n^{-\eps}$.
When the current interval has nominal mass at least a sufficiently large constant times $\delta$, an unsuccessful median query reduces its true mass by a constant factor. Since every median query reduces the remaining nominal mass by at least half, after $h=\lceil\eps\log n\rceil=\lceil\log(1/\delta)\rceil$ unsuccessful queries this mass is at most $\delta$. Then the CDF guarantee offers little guidance for further median queries, so we switch to binary search. Targets reaching this phase have $p_t=O(\delta)$, making $\log(1/p_t)$ large enough to pay for the additional $O(\log n)$ queries within a factor $O(1/\eps)$. Averaging over targets charges this cost to the entropy of $p$. The following theorem
 gives the precise argument; see its proof in \Cref{app:coarse-greedy}. \Cref{sec:information-limits} proves a matching lower bound.%

\begin{restatable}[Distributionally robust search from approximate CDFs]{theorem}{coarseGreedyTheorem}
\label{thm:CoarseGreedy}
Let $n\ge2$, $\varepsilon\in(0,1]$, and let
$U\subseteq\Delta_n$ be a nonempty uncertainty set.
Suppose there exists a nominal distribution $\pi\in\Delta_n$ such that
$
U\subseteq B_K(\pi,n^{-\varepsilon}).
$
Then the strategy $S$ obtained by running \textsc{Coarse Greedy}
with nominal distribution~$\pi$ satisfies
\[
\sup_{p\in U}
\frac{\operatorname{cost}_p(S)}{\operatorname{OPT}(p)}
\le O\left(\frac{1}{\varepsilon}\right).
\]
Consequently, $\alpha(\mathcal{U})=O(1/\eps)$ whenever every uncertainty set $U\in\mathcal{U}$ is contained in a Kolmogorov ball of radius $n^{-\eps}$. This holds in particular for the collection $\mathcal{U}_{K,n^{-\eps}}$ of all such balls.
\end{restatable}

\section{Information-Theoretic Limits of Uncertainty Sets}
\label{sec:information-limits}

We quantify the tradeoff between distributional information and search performance. First, we show that any sketch using at most $\frac12 n^\eps\log n$ bits has approximation ratio $\Omega(1/\eps)$ (\Cref{thm:informationLowerBound}). We extend this bound to arbitrary uncertainty collections through their covering number (\Cref{cor:Coveringlowerbound}). The bound is essentially tight: $n^\eps$ quantiles use $O(n^\eps\log n)$ bits and yield an $O(1/\eps)$-approximation via \textsc{Coarse Greedy} (\Cref{cor:quantileUpperBound}). Finally, we establish matching bounds for Kolmogorov balls (\Cref{thm:kolmogorovBallLowerBound}), and a sample-complexity guarantee appears in \Cref{subsec:sample-complexity}. %

\subsection{A Bit-Complexity Lower Bound}
\label{subsec:bit-lower-bound}

We first consider an uncertainty collection $\mathcal{U}_{\sigma}$
induced by a sketch $\sigma:\Delta_n\to\mathcal{X}$, where
$\mathcal{X}=\sigma(\Delta_n)$ is finite.
The sketch can be encoded with $k$ bits if $|\mathcal{X}|\le2^k$.
Equivalently, its uncertainty collection partitions $\Delta_n$ into
at most $2^k$ sets. We identify the messages with their binary encodings
in $\{0,1\}^k$.

The proof constructs a large family of distributions whose support points are spread over many intervals. If an encoding uses too few bits, then for some uncertainty set $U\in \mathcal{U}_{\sigma}$, the algorithm still has substantial uncertainty (measured as entropy) about the support point inside a large number of intervals. This residual uncertainty forces a large expected search cost.

We use standard entropy notation and properties; see Cover and Thomas~\cite[Chapter~2]{cover_elements_2006}. 
For a finite random variable $X$ with probabilities $p_x$, its Shannon entropy is $H(X):=-\sum_x p_x\log p_x$, where $0\log 0:=0$.
The joint entropy $H(X,Y)$ is the entropy of the pair $(X,Y)$.
We write $H(Y\mid X=x)$ for the entropy of the conditional distribution of $Y$ given $X=x$, and consider
\[
H(Y\mid X):=\sum_x\Pr(X=x)H(Y\mid X=x).
\]
We use the bound $H(X)\le\log|\operatorname{supp}(X)|$, the chain rule $H(X,Y)=H(X)+H(Y\mid X)$
and subadditivity $
H(Y_1,\ldots,Y_q)\le\sum_{j=1}^q H(Y_j),$
with equality when the $Y_j$ are independent.
Subadditivity also applies to the conditional distribution given $X=x$.

\begin{theorem}[Information-theoretic lower bound]
\label{thm:informationLowerBound}
For every fixed $\eps\in(0,1/4)$, there exists $n_0(\eps)$
such that the following holds for all $n\ge n_0(\eps)$.
Consider a sketch $\sigma$ that can be encoded with at most
$\frac12 n^\eps\log n$ bits. Then, for a universal constant $\gamma>0$, its uncertainty collection satisfies
\[
\alpha(\mathcal{U}_{\sigma})\ge\frac{\gamma}{\eps}.
\]
\end{theorem}

\begin{proof} We work with sufficiently large $n$, specifically $n\ge n_0(\eps):=\max\{2^{16},2^{1/\eps}\}$. 
Let $q:=\lceil n^\eps\rceil$, and partition $[n]_0$ into $q$ consecutive intervals $I_1,\ldots,I_q$ whose sizes differ by at most one.
Then $q\le n$ and $\log\lfloor n/q\rfloor\ge(1-\eps)\log n-2$, since $q\le2n^\eps$ and $\lfloor n/q\rfloor\ge n/(2q)$.

Our hard distributions have one support point in each interval, each carrying probability $1/q$. For $y\in I_1\times\cdots\times I_q$, denote the corresponding distribution by $p_y$. Choose $Y=(Y_1,\ldots,Y_q)$ by drawing each $Y_j$ independently and uniformly from $I_j$. Let $p_Y$ be the resulting random distribution and $X:=\sigma(p_Y)$ its (random) encoded message. The entropy of the support positions exceeds the sketch's bit budget, leaving substantial uncertainty even after observing the message. Indeed, since the $Y_j$ are independent and uniform on $I_j$, we have $H(Y)=\sum_j H(Y_j)=\sum_j\log|I_j|\geq q\log\lfloor(n/q)\rfloor$. The sketch uses at most $\frac12 n^\eps\log n$ bits, so $H(X)\le\log|\operatorname{supp}(X)|\le\frac12 n^\eps\log n$. Since $X$ is~determined~by~$Y$,
\[
H(Y\mid X)=H(X,Y)-H(X) = H(Y)-H(X)
\ge q\bigl((1-\eps)\log n-2\bigr)-\frac12 n^\eps\log n
\ge\frac18 q\log n,
\]
where we used $\eps<1/4$, $n^\eps\le q$, and $\log n\ge16$. Since $H(Y)\le q\log n$, observing the message can only remove a constant fraction of the uncertainty about $Y$.

Hence some message $x$ of positive probability leaves uncertainty $H(Y\mid X=x)\ge\frac18 q\log n$. Fix this message $x$. Any algorithm observing only the encoding must use the same search strategy for every distribution whose encoding is $x$; denote this strategy by $\mathcal{S}$. To translate this uncertainty into search cost, consider the average distribution $p':=\EE[p_Y\mid X=x]$, so $p'_t=q^{-1}\Pr(Y_j=t\mid X=x)$ for $t\in I_j$. Since each block has probability $1/q$ under $p'$, the entropy chain rule gives
\[
H(p')=\log q+\frac1q\sum_{j=1}^q H(Y_j\mid X=x)
\ge\frac1q H(Y\mid X=x)
\ge\frac18\log n.
\]
Here we used subadditivity and dropped the nonnegative term $\log q$.
By linearity of search cost in the target distribution and the entropy lower bound~\cite{mehlhornNearlyOptimalBinary1975},
\[
\EE[\operatorname{cost}_{p_Y}(\mathcal{S})\mid X=x]
=\operatorname{cost}_{p'}(\mathcal{S})
=\Omega(H(p'))
=\Omega(\log n).
\]
Thus, some distribution $p_y$ with $\sigma(p_y)=x$ incurs this cost for search strategy $\mathcal{S}$.

Knowing $p_y$, however, permits standard binary search among its $q=\lceil n^\eps\rceil$ support points, identifying every positive-probability target in $O(\log q)$ queries. Hence, $\opt(p_y)=O(\log q)$. Our choice $n\ge2^{1/\eps}$ ensures $\eps\log n\ge1$, so $\log q\le\eps\log n+1=O(\eps\log n) $. Thus, the approximation ratio on this input is $\Omega(\log n)/O(\eps\log n)=\Omega(1/\eps)$.
\end{proof}

\noindent\emph{Remark.} The lower bound also holds for randomized search strategies, even when the sketch is randomized; see~\Cref{sec:randomized-lower-bounds}.

We next extend the lower bound from sketches to arbitrary uncertainty collections. Let $\beta(\mathcal{U})$ be the minimum number of sets from $\mathcal{U}$ needed to cover $\Delta_n$, with $\beta(\mathcal{U})=\infty$ if no finite cover exists. A small cover induces a short encoding: assign each distribution the index of a covering set containing it. Applying \Cref{thm:informationLowerBound} to this encoding yields the following result.

\begin{restatable}{corollary}{coveringLowerBound}\label{cor:Coveringlowerbound}
For every fixed $\eps\in(0,1/4)$ and sufficiently large $n\ge n_0(\eps)$, any uncertainty collection $\mathcal{U}$ satisfying $\log\beta(\mathcal{U})\le\frac12 n^\eps\log n$ has approximability factor $\alpha(\mathcal{U})\ge\gamma/\eps$, where $\gamma>0$ is a constant.
\end{restatable}

\Cref{cor:Coveringlowerbound} applies even when the uncertainty sets overlap and needs no further assumption. To prove a lower bound, it suffices to construct a sufficiently small cover of the simplex by sets from the collection, without constructing a new family of hard inputs; see details in \Cref{app:information-limits-lower-bounds}.%

\subsection{Matching Bounds: Quantile Sketches and Kolmogorov Balls}
\label{subsec:quantile-upper-bound}

We show that the information-theoretic lower bound is tight %
by constructing quantile sketches, and establish tight bounds for the approximability of Kolmogorov balls. Proofs appear in \Cref{app:information-limits-upper-bounds}.

For an integer $q\ge 1$, define the $\ell$-th $q$-quantile of a distribution
$p\in\Delta_n$ as
\[
x_\ell:=\min\left\{t\in[n]_0:P(t)\ge \frac{\ell}{q}\right\},
\qquad \ell=1,\ldots,q,
\]
where $P$ is the CDF of $p$. Each quantile belongs to $[n]_0$ and can be stored using $\lceil\log n\rceil$ bits.

\begin{restatable}[Quantile upper bound]{theorem}{quantileUpperBound}
\label{cor:quantileUpperBound}
For every integer $q\ge1$, the $q$-quantiles of a distribution $p$ suffice to construct a distribution $\pi$ with $d_K(p,\pi)\le1/q$. Consequently, for $\eps\in(0,1)$, choosing $q=\lceil n^\eps\rceil$ gives a sketch of $O(n^\eps\log n)$ bits that guarantees an $O(1/\eps)$-approximation via \textsc{Coarse~Greedy}.
\end{restatable}

Thus, the $\Omega(1/\eps)$ lower bound is matched by the quantile upper bound up to constant factors in both the number of bits and the approximation ratio.
The construction places mass $1/q$ at each quantile value, adding masses when values repeat. Its CDF satisfies $\Pi(t)=\lfloor qP(t)\rfloor/q$, so rounding introduces error less than $1/q$.

We next characterize how the approximability of Kolmogorov balls depends on their radius. 
\begin{restatable}[Approximability of Kolmogorov balls]{theorem}{kolmogorovBallBounds}
\label{thm:kolmogorovBallLowerBound}
For every $n\ge2$ and $1/n\leq \delta \leq 1/2$,
\[
\alpha(\mathcal{U}_{K,\delta})
=\Theta\!\left(\frac{\log n}{\log(1/\delta)}\right).
\]
\end{restatable}

In particular, balls of radius $n^{-\eps}$ have approximability factor $\Theta(1/\eps)$ for $\eps\in(0,1]$ and $n\ge2^{1/\eps}$; by monotonicity, the lower bound also holds for larger radii.

The upper bound follows from \textsc{Coarse Greedy}. For the lower bound, we apply \Cref{cor:Coveringlowerbound}. %

Let $\mathcal{Q}_q$ be the set of distributions obtained by placing mass $1/q$ at each entry of a tuple $(x_1,\ldots,x_q)\in[n]_0^q$, adding masses when entries repeat; this is the construction from \Cref{cor:quantileUpperBound}, applied to arbitrary tuples. There are at most $n^q$ such distributions, and by \Cref{cor:quantileUpperBound} their Kolmogorov balls of radius $1/q$ cover $\Delta_n$.
Taking $q=\lceil1/\delta\rceil$ gives
\begin{equation}\label{eq:kolmogorov-cover}
\log\beta(\mathcal{U}_{K,\delta})
\le \lceil1/\delta\rceil\log n.
\end{equation}
Applying \Cref{cor:Coveringlowerbound} with a suitable parameter turns this cover-size estimate into the claimed lower bound, without requiring a new construction of hard distributions.

\section{Moment-Based Sketches}%
\label{sec:LB}

In this section, we study sketches based on moments. For $p\in\Delta_n$, the $i$-th moment is $\mu_i(p):=\sum_{t=0}^{n-1}p_t\,t^i$, and the $k$-moment sketch is $\sigma_k(p):=(\mu_0(p),\ldots,\mu_k(p))$. Our main result is negative: for $k\le n/(C'\log n)$, with $C'$ a suitable constant, every $k$-moment sketch has approximability factor $\Omega(\log\frac{n}{k\log n})$. In particular, for $k=n^{1-\delta}$ with constant $\delta>0$, the factor is $\Omega(\log n)$, which binary search already achieves without any information. This contrasts sharply with the $O(1/\eps)$ guarantee obtainable with $\tilde{O}(n^{\varepsilon})$ linear measurements in \Cref{sec:linear-sketches}. The proof proceeds in two stages. In \Cref{subsec:two-distributions}, we construct two distributions with identical moments but well-separated peaks, and extend this to many moment-matching distributions concentrated at different locations. \Cref{subsec:Moment-lowerbound-search} then turns this family into a search lower bound. Finally, \Cref{subsec:moment-positive} complements these negative results with structural conditions under which moments are informative, namely, bounded point masses and small support.

Here we present only the main ideas; the complete version is in \Cref{app:LB}.

\subsection{Distributions with Separated Peaks}
\label{subsec:two-distributions}

We start with a negative result: the next proposition shows that, even with the first $k$ moments fixed, two distributions can have most of their mass concentrated at two different positions, at distance of order $n/k$ from each other. Later, we will use this construction for an actual lower bound for our search problem.

\begin{restatable}[Two distributions with matching moments and separated peaks]{proposition}{twoDistros}\label{prop:2Distros}
There is an absolute constant $C>0$ such that the following holds. Let $n\ge3$ and $1\le k\le (n-1)/2$ be integers. For any two integers $m,m_0\in [\frac{n-1}{4},\frac{3(n-1)}{4}]$, there exist two distributions $p,p'$ on $[n]_0$ and a value $\gamma\le C\,\frac{k}{n-1}|m-m_0|$ such that
\[p_m,\, p'_{m_0}\ge1-\gamma
 \quad\text{and }\quad \mu_i(p)=\mu_i(p') \text{ for all }i=0,\ldots,k.\]
\end{restatable}

\begin{figure}[t]
  \centering
  \includegraphics[width=0.6\linewidth]{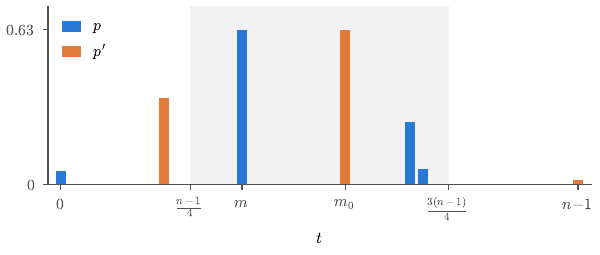}
  \caption{The construction in \Cref{prop:2Distros} for $n=41$, $k=4$, $m=14$, and $m_0=22$: the distributions $p$ and $p'$ have the same first $k$ moments. Shading marks $[\frac{n-1}4,\frac{3(n-1)}4]$.}
  \label{2Distros}
\end{figure}

See \Cref{2Distros} for an example of this proposition. The proposition implies that matching only the first $k$ moments does not determine where most of the mass lies. Indeed, we can select positions $m,m_0$ in the middle section of $[n]_0$ with $|m-m_0|=\Theta(n/k)$, while keeping $\gamma\le 1/4$ and ensuring the size of the peaks is at least $3/4$. %

The proof of the proposition proceeds in four steps. We first rescale the ground set to a uniform grid in $[-1,1]$. We then reduce the construction to finding a signed correction vector with small $\ell_1$-norm. This smallest norm is the value of a linear program, whose dual asks for the largest increment $q(\tau_m)-q(\tau_{m_0})$ of a polynomial $q$ of degree $k$ that is bounded by $1$ on the grid. Finally, we bound this increment using the classical Bernstein's inequality~\cite[Chapter~2]{rivlin1990} combined with a recent result by Rakhmanov~\cite{rakhmanov2007}.%

\paragraph{Step 1: Rescaling the ground set.}

Let $x(t):=\frac{2t-(n-1)}{n-1}$ be the affine map sending $[0,n-1]$ onto $[-1,1]$, and let $\tau_t:=x(t)$ for $t\in[n]_0$. Denote by $\mathcal{T}:=\{\tau_0,\ldots,\tau_{n-1}\}$ the corresponding scaled grid. Since $x$ is affine, moments are preserved: by \Cref{lm:rescaling} in the appendix, a vector $a\in\mathbb{R}^n$ satisfies $\sum_{t=0}^{n-1}a_t\,t^i=0$ for all $i=0,\ldots,k$ if and only if $\sum_{t=0}^{n-1}a_t\,\tau_t^i=0$ for all $i=0,\ldots,k$. Hence, two distributions $p,p'$ on $[n]_0$ have the same first $k$ moments if and only if $\sum_{t=0}^{n-1}p_t\tau_t^i=\sum_{t=0}^{n-1}p'_t\tau_t^i$ for all $i=0,\ldots,k$.

\paragraph{Step 2: Reduction to a signed correction.}

For $v\in[n]_0$, let $\delta_v\in\mathbb{R}^n$ denote the unit point mass at $v$, i.e., $(\delta_v)_t=1$ if $t=v$ and $(\delta_v)_t=0$ otherwise. For peak positions $m,m_0\in[n]_0$, consider the linear system
\begin{equation}\label{eq:Vandermondsystem}
\sum_{t=0}^{n-1} a_t \tau_t^i = \tau_m^i-\tau_{m_0}^i \qquad \text{for all }i=0,\ldots,k,
\end{equation}
in the unknown $a\in\mathbb{R}^n$. Let $a^+,a^-\ge0$ be the positive and negative parts of $a$, so that $a=a^+-a^-$. Then $a$ solves~\Cref{eq:Vandermondsystem} if and only if the nonnegative vectors $\delta_m+a^-$ and $\delta_{m_0}+a^+$ have the same first $k$ moments. In particular, by the equation for $i=0$, both vectors have the same total mass $1+\frac12\lVert a\rVert_1$, and normalizing them yields the following lemma.

\begin{restatable}[Signed corrections and distributions]{lemma}{signedVectorReduction}\label{lm:signedVectorReduction}
Let $m,m_0\in[n]_0$. If $a$ solves~\Cref{eq:Vandermondsystem}, then there are two distributions $p,p'$ on $[n]_0$ with the same first $k$ moments and $p_m,\,p'_{m_0}\ge1-\frac12\lVert a\rVert_1$.
\end{restatable}

\Cref{lm:signedVectorReduction} shows that, for given peak positions $m,m_0$, the question is how small the $\ell_1$-norm of a solution of~\Cref{eq:Vandermondsystem} can be. We therefore define
\begin{equation}\label{eq:gammastar}
\gamma^*(m,m_0):=\inf\bigl\{\lVert a\rVert_1 \,:\, a\in\mathbb{R}^{n} \text{ satisfies~\Cref{eq:Vandermondsystem}}\bigr\}.
\end{equation}
It thus remains to bound $\gamma^*(m,m_0)$, which we do in the next two steps. %

\paragraph{Step 3: The dual of the correction problem.}

Writing $a=a^+-a^-$ with $a^+,a^-\ge0$ as above, the value $\gamma^*(m,m_0)$ is the optimum of a linear program, which we call the \emph{correction LP}. Its dual is 
\[
        \text{\rm [D]}\qquad
        \max\left\lbrace \sum_{i=0}^k \lambda_i \bigl(\tau_m^i-\tau_{m_0}^i\bigr) \,:\,
        -1\le \sum_{i=0}^k \lambda_i \tau_t^i \le 1 \quad \text{for all } t=0,\ldots,n-1\right\rbrace.
\]
 We can interpret a dual solution $(\lambda_0,\ldots,\lambda_k)$ as the polynomial $q(x)=\sum_{i=0}^k\lambda_i x^i$ of degree at most~$k$. The constraints say that $|q(\tau_t)|\le1$ for all $t\in[n]_0$. We call such a $q$ a \emph{feasible polynomial}. The objective is $q(\tau_m)-q(\tau_{m_0})$. The correction LP is feasible (e.g., $a^+=\delta_m$, $a^-=\delta_{m_0}$) and its objective is bounded below by $0$, so its optimum is attained, and by strong linear programming duality it coincides with the optimum of [D]. Together with the observation above that this LP computes $\gamma^*(m,m_0)$, this gives the following lemma.

\begin{restatable}[Dual of the signed correction]{lemma}{dualCorrection}\label{lem:dualCorrection}
For all $m,m_0\in[n]_0$,
\[
  \gamma^*(m,m_0)
  =\max\bigl\{\,q(\tau_m)-q(\tau_{m_0})\ :\ q \text{ a polynomial with } \deg q\le k \text{ and }
  |q(\tau_t)|\le 1\ \text{for all } t\in[n]_0\,\bigr\}.
\]
Moreover, the infimum in~\Cref{eq:gammastar} is attained.
\end{restatable}

With this lemma, we get that to upper bound $\gamma^*(m,m_0)$, it suffices to upper bound $q(\tau_m)-q(\tau_{m_0})$ for every feasible polynomial. %

\paragraph{Step 4: Bounding the increments of feasible polynomials.}

To bound $q(\tau_m)-q(\tau_{m_0})$, we bound the derivative of $q$ and use the mean value theorem. We use the following classical inequality for real polynomials $q$ with $\deg q\le k$; see, e.g.,~\cite[Chapter~2]{rivlin1990}. Writing $\lVert q\rVert:=\max_{x\in[-1,1]}|q(x)|$,
\begin{align}
  \text{(Bernstein)}\qquad &|q'(x)|\;\le\;\frac{k}{\sqrt{1-x^{2}}}\,\lVert q\rVert
  \qquad \text{for all } x\in(-1,1). \label{eq:bernstein}
\end{align}
Applying~\Cref{eq:bernstein} to $x\mapsto q(\rho x)$ shows that, for every $\rho>0$,
\begin{equation}\label{eq:bernsteinScaled}
  |q(x)|\le B \text{ for all } x\in[-\rho,\rho]
  \quad\Longrightarrow\quad
  |q'(x)|\le\frac{k\,B}{\sqrt{\rho^2-x^2}}\quad\text{for all } x\in(-\rho,\rho).
\end{equation}

The constraint in \Cref{lem:dualCorrection} bounds $q$ on the grid $\mathcal{T}$ only, whereas~\Cref{eq:bernstein} requires a bound on an interval. To transfer the bound from the grid to (the central part of) the interval, we use a landmark theorem of Rakhmanov~\cite[Theorem~1]{rakhmanov2007}. Adapted to our setting, this result implies the following.

\begin{restatable}[From the grid to the central interval]{lemma}{gridToCentralInterval}\label{lem:gridCentral}
Let $q$ be a polynomial with $\deg q\le k\le (n-1)/2$ and $|q(\tau_t)|\le1$ for all $t\in[n]_0$. Then $|q(x)|\le C_1$ for all $x\in[-3/4,3/4]$ for some absolute constant~$C_1$.
\end{restatable}

We can now bound the value of the correction problem.

\begin{restatable}{lemma}{cheapCorrection}%
  \label{lem:cheapCorrection}
Let $1\le k\le (n-1)/2$ and let $m,m_0\in[n]_0\cap[\frac {n-1}4,\frac{3(n-1)}4]$. Then
\[
  \gamma^*(m,m_0)\;\le\;\frac{8C_1}{\sqrt5}\,\frac{k\,|m-m_0|}{n-1},
\]
where $C_1$ is the constant of \Cref{lem:gridCentral}.
\end{restatable}

\begin{proof}
Let $q$ be an optimal solution of the maximization problem in \Cref{lem:dualCorrection}, so that $\gamma^*(m,m_0)=q(\tau_m)-q(\tau_{m_0})$. Recall that $\tau_m,\tau_{m_0}\in[-\frac12,\frac12]$. By \Cref{lem:gridCentral} and \Cref{eq:bernsteinScaled} with $\rho=\frac34$, we obtain that for all $x\in[-\frac12,\frac12]$,
\[
|q'(x)|\le\frac{k\,C_1}{\sqrt{\frac9{16}-\frac14}}=\frac{4C_1}{\sqrt5}\,k.
\]
By the mean value theorem, $
\gamma^*(m,m_0)=q(\tau_m)-q(\tau_{m_0})\le\frac{4C_1}{\sqrt5}\,k\,|\tau_m-\tau_{m_0}|=\frac{8C_1}{\sqrt5}\,\frac{k\,|m-m_0|}{n-1}.
$
\end{proof}

Finally, \Cref{prop:2Distros} follows by combining \Cref{lem:cheapCorrection} and \Cref{lm:signedVectorReduction}.

We now generalize the construction from two distributions to several distributions with the same first $k$ moments. To get distributions with several peaks at $m_0,m_1,\ldots,m_R$, we construct vectors $a^{(s)}\in\mathbb{R}^{n}$ optimizing~\Cref{eq:gammastar} for the pair $(m_s,m_0)$, and then combine them appropriately. The cost is obtaining smaller peaks.

\begin{restatable}[Several distributions with matching moments]{proposition}{manyDistrosSameMoments}\label{prop:manyDistrosSameMoments}
Let $n\ge3$ and $1\le k\le (n-1)/2$, and let $
m_0,m_1,\ldots,m_R \in \left[\frac {n-1}4,\frac{3(n-1)}4 \right]\cap[n]_0.$ Let $C$ be the constant of \Cref{prop:2Distros}. Then there exist distributions $p^{(0)},p^{(1)},\ldots,p^{(R)}$ on $[n]_0$ with the same first $k$ moments and
\[
p^{(r)}_{m_r}\ge 1-\gamma \qquad \text{for all }r=0,\ldots,R,
\qquad\text{where}\qquad
\gamma \le C\,\frac{k}{n-1}\sum_{s=1}^R |m_s-m_0|.\]
\end{restatable}

\begin{proof}%
We use the signed-correction construction from the proof of \Cref{prop:2Distros}. For $s=1,\ldots,R$, let $a^{(s)}\in\mathbb{R}^{n}$ be a minimizer of~\Cref{eq:gammastar} for the pair $(m_s,m_0)$. By \Cref{lem:cheapCorrection},
\[
\lVert a^{(s)}\rVert_1\le 2C\,\frac k{n-1}|m_s-m_0| \qquad\text{for all } s=1,\ldots,R.
\]
Write $a^{(s)}=(a^{(s)})^+-(a^{(s)})^-$, where $(a^{(s)})^{\pm}\ge0$ are the positive and negative parts of $a^{(s)}$. %

Define $B:=\sum_{s=1}^R (a^{(s)})^+$ and $Z:=1+\lVert B\rVert_1$, and let
\[
p^{(0)} := \frac{\delta_{m_0}+B}{Z}
\qquad\text{and}\qquad
p^{(r)} := \frac{\delta_{m_r} + (a^{(r)})^- + \sum_{s\neq r}(a^{(s)})^+ }{Z}
\quad\text{for } r=1,\ldots,R.
\]

All these vectors are nonnegative. Since the constraint for $i=0$ in~\Cref{eq:Vandermondsystem} gives $\lVert(a^{(s)})^+\rVert_1=\lVert(a^{(s)})^-\rVert_1=\frac12\lVert a^{(s)}\rVert_1$, each numerator has total mass $Z$, so $p^{(0)},\ldots,p^{(R)}$ are probability distributions on $[n]_0$. For $r\ge1$, the numerators of $p^{(0)}$ and $p^{(r)}$ differ by $a^{(r)}+\delta_{m_0}-\delta_{m_r}$, so by \Cref{lm:rescaling} all these distributions have the same first $k$ moments. Finally, each peak satisfies $p^{(r)}_{m_r}\ge\frac1Z\ge1-\lVert B\rVert_1\ge 1- C\,\frac k{n-1}\sum_{s=1}^R |m_s-m_0|$.
\end{proof}

\subsection{A Search Lower Bound from Moment Matching}\label{subsec:Moment-lowerbound-search}

Based on the construction in ~\Cref{prop:manyDistrosSameMoments}, we can show the main result of this section.

\begin{restatable}[Exact moments can be useless for search]{theorem}{momentSearchSeparation}
\label{thm:momentSearchSeparation}
There is an absolute constant $C'>0$ such that, for all sufficiently large $n$ and every integer $1\le k\le n/(C'\log n)$, the uncertainty collection $\mathcal U_{\sigma_k}$ induced by the moment sketch $\sigma_k(p):=(\mu_0(p),\ldots,\mu_k(p))$ satisfies
\[
\alpha(\mathcal U_{\sigma_k})=\Omega\Bigl(\log\frac{n}{k\log n}\Bigr).
\]
In particular, the bound is $\Omega(\log\frac{n}{k})$ for $k\le\frac{n}{\log^2 n}$, and $\Omega(\log n)$ for $k\le n^{1-\delta}$ with constant $\delta\in(0,1]$.
\end{restatable}

The lower bound also holds for randomized search strategies; see~\Cref{sec:randomized-lower-bounds}.

\begin{proof}[Proof idea.]
Apply \Cref{prop:manyDistrosSameMoments} with $R=\Theta\bigl(\sqrt{n/(k\log n)}\bigr)$ consecutive peaks $m_r:=m_0+r$, $r=0,\ldots,R$, around the middle of $[n]_0$. This yields distributions $p^{(0)},\ldots,p^{(R)}$ with the same first $k$ moments and peaks $p^{(r)}_{m_r}\ge1-\gamma$, where $\gamma=O(kR^2/n)\le 1/(4\log n)$ for a suitable choice of the constant in $R$. In particular, all of them lie in a common uncertainty set $U\in\mathcal U_{\sigma_k}$.

Fix a deterministic search strategy $T$ for $U$. Interpreting $T$ as a ternary tree, Kraft's inequality~\cite[Theorem~5.2.1]{cover_elements_2006} implies that $T$ uses at least $\log_3(R+1)$ queries to identify some peak $m_r$, so its expected cost on $p^{(r)}$ is $\Omega(\log(R+1))$. In contrast, a strategy that knows $p^{(r)}$ first queries $m_r$ and, if needed, finishes with binary search, so $\opt(p^{(r)})\le1+\gamma\lceil\log_2 n\rceil=O(1)$. Hence every strategy for $U$ has approximation ratio $\Omega(\log(R+1))=\Omega\bigl(\log\frac{n}{k\log n}\bigr)$ on some distribution in $U$.
\end{proof}

\subsection{Approximate and Exact Recovery from Moments}
\label{subsec:moment-positive}

We also give self-contained proofs of recovery guarantees under bounded-atom and small-support assumptions. Both require only that one distribution consistent with the observed moments satisfies the stated condition.

Both arguments operate on the same principle: we use an LP to compute the maximum or minimum values attainable by the CDF, and then create a dual certificate using polynomial interpolation techniques. It is worth mentioning that the LP we consider is the finite-dimensional version of the semi-infinite linear program considered by Bertsimas and Popescu~\cite{bertsimasPopescu2005} and Bertsimas and Sethuraman~\cite{bertsimas_moment_2000}, but also considered implicitly in the classical literature of moments; see, e.g., the monograph by Karlin and Studden~\cite{karlin1966tchebycheff}. 

\begin{figure}[t]
  \centering
  \includegraphics[width=0.6\linewidth]{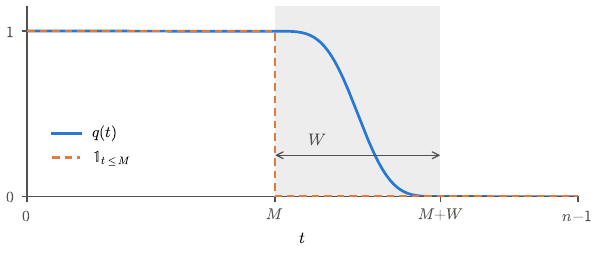}
  \caption{
  Illustration of the smoothed-step construction in \Cref{lem:cert}. The polynomial~$q$ approximates the indicator $\mathbbm{1}_{t\le M}$, with the shaded interval $[M,M+W]$ marking the transition~window.}
  \label{smoothedStep}
\end{figure}

Fix the moments $\bar\mu_0,\ldots,\bar\mu_k$ of the (unknown) true distribution, and let $\pi$ be a nominal distribution with these moments, that is, $\bar\mu_i=\sum_{t=0}^{n-1} t^i \pi_t$ for $i=0,\ldots,k$; such a $\pi$ can be computed from the moments by solving a linear feasibility problem. For $M\in[n-1]_0$, the largest possible CDF value at $M$, among distributions with matching moments, is given by the linear program [P$'$] below; its dual is [D$'$].
\begin{alignat*}{2}
        \text{\rm [P$'$]}\quad \max\quad & \sum_{t=0}^M p_t
        &\hspace{2em} \text{\rm [D$'$]}\quad \min\quad & \sum_{i=0}^k \lambda_i \bar\mu_i\\
        \text{s.t.}\quad & \sum_{t=0}^{n-1} t^i p_t=\bar\mu_i \quad \text{for } i=0,\ldots,k,
        & \text{s.t.}\quad & \sum_{i=0}^k \lambda_i t^i \ge 1 \quad \text{for } t=0,\ldots,M,\\
        & p_t\ge 0 \quad \text{for } t=0,\ldots,n-1,
        & & \sum_{i=0}^k \lambda_i t^i \ge 0 \quad \text{for } t=M+1,\ldots,n-1.
\end{alignat*}
Hence, as for [D] in \Cref{subsec:two-distributions}, a dual solution $(\lambda_0,\ldots,\lambda_k)$ can be identified with a polynomial
$q(x)=\sum_{i=0}^k\lambda_i x^i$ of degree at most $k$. Thus, a feasible dual solution is a polynomial satisfying
\[
q(t)\ge 1 \quad \text{for } t\in \{0,1,\ldots,M\},
\qquad \text{and }
q(t)\ge 0 \quad \text{for } t\in \{M+1,\ldots,n-1\}.
\]
Moreover, we can rewrite the dual objective as
\[
\sum_{i=0}^k \lambda_i \bar\mu_i
= \sum_{i=0}^k \lambda_i \sum_{t=0}^{n-1} t^i \pi_t
= \sum_{t=0}^{n-1} \pi_t \sum_{i=0}^k \lambda_i t^i
= \sum_{t=0}^{n-1} \pi_t\, q(t).
\]
If some polynomial of degree $k$ coincided with $\mathbbm{1}_{t\le M}$ on $[n]_0$, it would be a dual solution of value exactly $\sum_{t=0}^{M}\pi_t$, implying $\sum_{t=0}^{M}p_t \le \sum_{t=0}^{M}\pi_t$; a symmetric argument would then give equality. For $k<n-1$, such a polynomial does not exist, and the task becomes approximating $\mathbbm{1}_{t\le M}$ by a polynomial of degree $k$. Properties of Chebyshev polynomials allow us to do so, which yields the following lemma; see \Cref{smoothedStep} for an illustration.

Bustamante, Mart\'inez-Cruz, and Quesada~\cite{bustamanteMQ2015} characterize best one-sided polynomial approximations to step functions in the continuous $L_1$ norm; our construction provides pointwise bounds yielding CDF recovery guarantees for bounded-atom distributions on a finite grid.

\begin{restatable}[One-sided smoothed step]{lemma}{smoothedStep}\label{lem:cert}
Let $n\ge3$, $k\ge16$, $M\in[0,n-1]$, and let $W\in \mathbb{Z}$ %
with
$96\,\frac{n}{k}\ln n \;\le\; W\;\le\;\frac{n-1}{2}$ and $M+W\le n-1.$ Then there is a polynomial $q$ with $\deg q\le k$ such that:
\begin{multicols}{2}
\begin{enumerate}[label=(\roman*)]
  \item $q(t)\ge0$ for all $t\in[0,n-1]$;
  \item $q(t)\ge1$ for all $t\in[0,M]$;
  \item $q(t)\le 1+n^{-2}$ for all $t\in[0,n-1]$;
  \item $q(t)\le n^{-2}$ for all $t\in[M+W,\,n-1]$.
\end{enumerate}
\end{multicols}
\end{restatable}

The next theorem quantifies how small atoms in a nominal
distribution limit the CDF ambiguity left by its first $k$ moments.
The bound applies to every distribution with the same moments,
without requiring the same atom bound for those distributions. The proof relies on the dual construction in \Cref{lem:cert} and weak duality.

\begin{restatable}[Moment recovery under bounded atoms]{theorem}{momentRecoveryBoundedAtoms}
\label{thm:momentRecoveryBoundedAtoms}
Let $16\le k\le n$, and let $\pi$ be any distribution on $[n]_0$ with $\|\pi\|_\infty\le L$. Then every distribution $p$ on $[n]_0$ with the same first $k$ moments as $\pi$ satisfies
\[
d_K(\pi,p)=O\Big(L\,\frac{n}{k}\log n\Big).
\]
\end{restatable}
For a related result on CDF recovery from matching moments when one distribution has a bounded density, see Krawtchouk~\cite{krawtchouk1932} and the discussion in Wu and Yang~\cite[Remark~2]{wuYang2020}.

Some dependency on $L$ in the theorem is necessary: for some $k=\Omega(n)$, \Cref{prop:2Distros} yields two distributions with $k$ matching moments, a peak of mass $\Omega(1)$, and Kolmogorov distance $\Omega(1)$. Hence, no bound independent of $L$ can tend to $0$, even with linearly many moments. In contrast, for $k=\Theta(n)$, our bound tends to $0$ whenever $L=o(1/\log n)$. %

\Cref{thm:momentRecoveryBoundedAtoms} gives a quantitative CDF-recovery bound. In the parameter regime where its bound guarantees $d_K(\pi,p)\le n^{-\eps}$, we also have $L=O(n^{-\eps})$. Every compatible distribution therefore satisfies $p_t\le\pi_t+2d_K(\pi,p)=O(n^{-\eps})$ for every $t$. Thus, $H(p)=\Omega(\eps\log n)$, and ordinary binary search already achieves an $O(1/\eps)$-approximation in this regime.

Finally, we recall that if a nominal distribution with the observed moments has support of size $\ell$, then $k=2\ell$ moments determine the true distribution exactly, and %
we can optimize without error. This is achieved by using polynomial interpolation that exactly matches $\mathbbm{1}_{t\le M}$ on the support of $\pi$.

This identifiability fact appears in Wu and Yang~\cite[Lemma~4(2)]{wuYang2020}; we include a proof in our LP-duality framework for completeness.

\begin{restatable}[Moment recovery for small support]{proposition}{momentRecoverySmallSupport}
\label{prop:momentRecoverySmallSupport}
Let $\pi$ be a distribution on $[n]_0$ with support size $\ell$, and let $k\ge2\ell$. Suppose that $p$ is a distribution on $[n]_0$ with the same first $k$ moments as $\pi$. Then~$p=\pi$.
\end{restatable}

The preceding results show that moments can be informative under strong structural assumptions: either some nominal distribution with the given moments has sufficiently small atoms, or one has very small support. As shown in \Cref{subsec:two-distributions,subsec:Moment-lowerbound-search}, without such assumptions the first $k$ moments may be compatible with distributions whose mass is concentrated at very different locations, which makes it impossible to search with a close-to-optimal algorithm.

\section{Linear Sketches}
\label{sec:linear-sketches}

The limitations of moments do not extend to all linear measurements.
A linear sketch is specified by a matrix $A\in\mathbb{R}^{k\times n}$
and reveals $Ap$. We count the number $k$ of exact measurements,
rather than their bit complexity; \Cref{app:linear-sketches} discusses finite-precision sketches and their relation to~\cite{gribelyuk_lifting_2025}. 
Let $\mathcal{U}_{\sigma_A}$
denote the uncertainty collection induced by $\sigma_A(p)=Ap$.

\begin{restatable}[Linear sketch lower bound]{theorem}{linearSketchLowerBound}
\label{thm:linearSketchlowerBound}
For every $k\ge2$ and every matrix $A\in\mathbb{R}^{k\times n}$,
\[
\alpha(\mathcal{U}_{\sigma_A})
=\Omega\left(\frac{\log n}{\log k}\right).
\]
\end{restatable}

The proof considers the uncertainty set containing the uniform
distribution. For any strategy chosen for this set, the uniform
distribution incurs cost $\Omega(\log n)$. Maximizing the strategy's
cost over the set yields an optimal extreme point supported on at
most $k+1$ positions. This distribution incurs at least the same
cost, but a strategy knowing it needs only $O(\log k)$ expected
queries.

Thus $n^{\Omega(\eps)}$ measurements are necessary for an
$O(1/\eps)$-approximation. The next result shows that
$\widetilde O(n^\eps)$ measurements suffice. In particular, the poor
performance of moments reflects the choice of measurements,
rather than linearity itself.

\begin{restatable}[Search from linear sketches]{corollary}{linearSketchSearchUpperBound}
\label{cor:searchFromDyadicJL}
For every $\eps\in(0,1)$, there is a matrix
$V\in\mathbb{R}^{k\times n}$ with
$k=O(n^\eps\log^4 n)$ such that
$\alpha(\mathcal{U}_{\sigma_V})=O(1/\eps)$.
\end{restatable}

For the upper bound, we adapt the dyadic decomposition used in
range-query and quantile sketches
\cite{gilbert2002summarize,cormode2005improved}, combining it with
near-orthogonal vectors obtained from the Johnson--Lindenstrauss
lemma~\cite{johnson1984extensions}.
We sketch interval masses separately at each dyadic level.
Near-orthogonality bounds each interval-mass estimation error by the inner-product error times the total mass at that level. Since this mass is one, the bound is independent of the number of intervals.
Every prefix is a union of $O(\log n)$ dyadic intervals, so their
estimates yield a uniform CDF approximation. The construction
achieves error $\delta$ using $O(\delta^{-2}\log^4 n)$ measurements.
Choosing any distribution consistent with the sketch gives a
valid nominal CDF with at most twice this error.
Taking $\delta=\tfrac12 n^{-\eps/2}$ and running \textsc{Coarse Greedy} with parameter $\eps/2$ and the resulting nominal distribution gives the search guarantee by \Cref{thm:CoarseGreedy}.
The full proofs and sketch construction appear in
\Cref{app:linear-sketches}.

\section{Concluding Remarks}

For the fundamental problem of binary search, we have identified sharp differences in the power of distributional information, comparing arbitrary bit encodings, moments, quantiles, approximate CDFs, and general linear sketches. Our results establish quantitative information requirements and show that the choice of sketch can matter far more than its size. Sample access provides another route: classical CDF concentration bounds imply that $O(n^\eps\log(2/\delta))$ independent samples suffice for an $O(1/\eps)$-approximation with probability at least $1-\delta$ (\Cref{subsec:sample-complexity}).

These results open a research direction: systematically comparing the power of distributional summaries of comparable size, with performance measured against an optimum that knows the true distribution. A natural next step is to investigate whether similarly strong separations between moments and other summaries hold for scheduling, pricing, or other stochastic optimization problems.

\clearpage
\appendix
\section*{Appendices}
\addcontentsline{toc}{section}{Appendices}

\crefalias{section}{appendix}
\crefalias{subsection}{subappendix}
\crefalias{subsubsection}{subsubappendix}

\section{Proof of the Coarse Greedy Guarantee
(\Cref{sec:model-coarse-greedy})}
\label{app:coarse-greedy}

\coarseGreedyTheorem*

\begin{proof}
Fix any $p\in U$ and put $\delta:=n^{-\varepsilon}$.
By assumption, $d_K(p,\pi)\le\delta$.
The CDF bound implies, for every target $t$,
\[
|p_t-\pi_t|
=
|(P(t)-\Pi(t))-(P(t-1)-\Pi(t-1))|
\le2\delta.
\]
Every unsuccessful median query reduces the remaining nominal mass by at least half.
Moreover, the algorithm makes at most
$h+\lceil\log n\rceil=O(\log n)$ queries on any target. If $\varepsilon\log n<4$, this worst-case bound is
$O(1/\varepsilon)$, and the claim follows from
$\operatorname{OPT}(p)\ge1$.
Hence assume $\varepsilon\log n\ge4$.

Fix a target $t$ with $p_t>0$. We distinguish whether its probability is large compared to $\delta$.

If $p_t>4\delta$, then $\pi_t\ge p_t-2\delta>p_t/2>2\delta$. This ensures that $t$ is identified within the first $h$ median queries, without reaching the binary-search fallback. Indeed, after $j$ unsuccessful median queries, the remaining interval
$I$ contains $t$ and satisfies $\pi_t\le\pi(I)\le2^{-j}$.
Hence $t$ cannot reach an interval of nominal mass zero or
survive all $h$ median queries, since $2^{-h}\le\delta<\pi_t$.

Let $r=q(S,t)$ be the number of queries before $t$ is identified, either by equality or by reaching a singleton.
After the first $r-1$ queries, the remaining interval contains $t$
and has nominal mass at most $2^{-(r-1)}$.
Hence $\pi_t\le2^{-(r-1)}$, which gives
$r\le1+\log(1/\pi_t)$.
Its query cost is therefore
\[
q(S,t)=r\le1+\log(1/\pi_t)
\in O(1+\log(1/p_t)),
\]
where the last bound uses $\pi_t>p_t/2$.

If instead $p_t\le4\delta$, the nominal probability may be zero, so we use the worst-case query bound, whether or not $t$ reaches binary search. Its small true probability makes $\log(1/p_t)$ large enough to pay for this cost: since $\eps\log n\ge4$,
\[
\log(1/p_t)\ge\eps\log n-2\ge\tfrac12\eps\log n.
\]
The worst-case query bound therefore yields
\[
q(S,t)
\in O(\log n)
=O\left(\frac1\varepsilon \log(1/p_t)\right).
\]

Combining the two cases and taking expectations over $t\sim p$ yields
\[
\operatorname{cost}_p(S)\le O\left(\frac{1+H(p)}{\eps}\right).
\]
Using the entropy lower bound
$H(p)=O(\opt(p))$~\cite{mehlhornNearlyOptimalBinary1975}
together with $\opt(p)\ge1$, we obtain
$
\operatorname{cost}_p(S)\le O(1/\eps)\opt(p).
$
Since $p\in U$ was arbitrary and $S$ depends only on $\pi$ and
$\eps$, this guarantee holds uniformly over $U$.

Choosing a nominal distribution as assumed for each $U\in\mathcal{U}$ gives $\alpha(\mathcal{U})=O(1/\eps)$.
\end{proof}

\section{Proofs for Information-Theoretic Limits (\Cref{sec:information-limits})}
\label{app:information-limits-proofs}

\subsection{Lower Bounds for General Uncertainty Collections}
\label{app:information-limits-lower-bounds}

For an uncertainty collection $\mathcal{U}$, let $\beta(\mathcal{U})$ be the minimum number of sets from $\mathcal{U}$ whose union is $\Delta_n$, and set $\beta(\mathcal{U})=\infty$ if no finite subcollection covers $\Delta_n$. The following generalization of \Cref{thm:informationLowerBound} applies even when the sets overlap and need not be convex or defined by a metric.

\coveringLowerBound*

\begin{proof}
Set $n_0(\eps):=\max\{2^{16},2^{1/\eps}\}$, as in the proof of \Cref{thm:informationLowerBound}.

Let $U_1,\ldots,U_{\beta(\mathcal{U})}\in\mathcal{U}$ be a minimum cover of $\Delta_n$. Consider the sets
\[
\tilde{U}_i = U_i\setminus \bigcup_{j<i} U_j \qquad \text{for all } i=1,\ldots,\beta(\mathcal{U}).
\]
Each $\tilde{U}_i$ is nonempty: otherwise $U_i$ would be contained
in the union of the preceding sets and could be removed from the
minimum cover.
Then $\widetilde{\mathcal{U}}:=\{\tilde{U}_1,\ldots,
\tilde{U}_{\beta(\mathcal{U})}\}$ defines a partition of $\Delta_n$.
The corresponding sketch $\sigma(p)=i$ for $p\in\tilde{U}_i$ has at most $\beta(\mathcal{U})$ messages, so its message entropy is at most $\log\beta(\mathcal{U})\le\frac12 n^\eps\log n$. The entropy argument in the proof of \Cref{thm:informationLowerBound} therefore gives $\alpha(\widetilde{\mathcal{U}})\ge\gamma/\eps$.
Since $\tilde{U}_i\subseteq U_i$, the definition of $\alpha$ gives
\begin{align*}
\alpha(\mathcal{U})
&\ge \sup_{1\le i\le\beta(\mathcal{U})}
  \inf_S\sup_{p\in U_i}
  \frac{\operatorname{cost}_p(S)}{\opt(p)}\\
&\ge \sup_{1\le i\le\beta(\mathcal{U})}
  \inf_S\sup_{p\in\tilde{U}_i}
  \frac{\operatorname{cost}_p(S)}{\opt(p)}
=\alpha(\widetilde{\mathcal{U}})\ge\frac{\gamma}{\eps}.  
\end{align*}
\end{proof}

\subsection{Quantile Upper Bounds and the Kolmogorov-Ball Characterization}
\label{app:information-limits-upper-bounds}

\quantileUpperBound*
\begin{proof}
Let $x_1,\ldots,x_q$ be the $q$-quantiles of $p$. Define $\pi$ by placing
mass $1/q$ at each quantile value, counting multiplicities. Thus, if $\Pi$
denotes the CDF of $\pi$, then
\[
\Pi(t)=\frac1q
\left|\left\{\ell\in\{1,\ldots,q\}:x_\ell\le t\right\}\right|.
\]
By the definition of $x_\ell$, we have $x_\ell\le t$ if and only if
$\ell/q\le P(t)$. It follows that
$\Pi(t)=\lfloor qP(t)\rfloor/q$. Therefore,
$0\le P(t)-\Pi(t)<1/q$ for every $t\in[n]_0$, and hence
$d_K(p,\pi)\le1/q$.

For $q=\lceil n^\eps\rceil$, the error is at most $n^{-\eps}$, and storing the quantiles requires $q\lceil\log n\rceil=O(n^\eps\log n)$ bits. The approximation guarantee follows from \Cref{thm:CoarseGreedy}.
\end{proof}

\kolmogorovBallBounds*
\begin{proof}
For the upper bound, set
$\eps:=\log(1/\delta)/\log n\in(0,1]$, so that $\delta=n^{-\eps}$.
\Cref{thm:CoarseGreedy} gives
\[
\alpha(\mathcal{U}_{K,\delta})
=O(1/\eps)
=O\!\left(\frac{\log n}{\log(1/\delta)}\right).
\]

For the lower bound, if $\eps\ge1/16$, then
$\alpha(\mathcal{U}_{K,\delta})\ge1=\Omega(1/\eps)$.
Otherwise, since $\eps\log n=\log(1/\delta)\ge1$, we have
$3\eps\in(0,1/4)$, $\log n\ge1/\eps>16$, and
$3\eps\log n\ge3$. Thus $n\ge n_0(3\eps)$ for the choice made
in the proof of \Cref{cor:Coveringlowerbound}.
By \Cref{eq:kolmogorov-cover} and $n^\eps=1/\delta\ge2$,
\[
\log\beta(\mathcal{U}_{K,\delta})
\le (n^\eps+1)\log n
\le 2n^\eps\log n
\le \frac12 n^{3\eps}\log n.
\]
Applying \Cref{cor:Coveringlowerbound} with parameter $3\eps$ yields
\[
\alpha(\mathcal{U}_{K,\delta})
\ge \frac{\gamma}{3\eps}
=\Omega\!\left(\frac{\log n}{\log(1/\delta)}\right).\qedhere 
\]
\end{proof}

\section{Moment-Based Sketches: Full Version (\Cref{sec:LB})}
\label{app:LB}

This appendix contains a complete version of \Cref{sec:LB}, including all proofs. All statements are restated with their numbering from the main text.

In \Cref{sec:information-limits}, we showed that a sketch
using $O(n^\eps\log n)$ bits suffices for an
$O(1/\eps)$-approximation. 
Moments, however, are exact real numbers, so the moment sketch has uncountably many
possible values and its uncertainty collection has infinite covering number. Hence the
bit-complexity lower bounds of \Cref{sec:information-limits} do not transfer to this setting. Moreover, the hard instances in the proof of \Cref{thm:informationLowerBound} are easy for moments: each has support of size $q=\lceil n^\eps\rceil$, so by \Cref{prop:momentRecoverySmallSupport} below, its first $2q=O(n^\eps)$ moments determine it exactly. Regardless, a natural question is whether $n^{\varepsilon}$ moments can be used to obtain an $O(1/\varepsilon)$-approximation. We emphasize that the $n^{\varepsilon}$ benchmark is natural in this context: in \Cref{sec:linear-sketches} we show that any linear sketch needs at least $n^{\Omega(\varepsilon)}$ linear measurements for an $O(1/\varepsilon)$ guarantee.

Our main result is negative: for $k\le n/(C'\log n)$, with $C'$ a suitable constant, every $k$-moment sketch has approximability factor $\Omega(\log\frac{n}{k\log n})$. In particular, for $k=n^{1-\delta}$ with constant $\delta>0$, the factor is $\Omega(\log n)$, which binary search already achieves without any information. This contrasts sharply with the $O(1/\eps)$ guarantee obtainable with $\tilde{O}(n^{\varepsilon})$ linear measurements in \Cref{sec:linear-sketches}. The proof proceeds in two stages. In \Cref{app:two-distributions-app}, we construct two distributions with identical moments but well-separated peaks, and in \Cref{app:many-distributions-app} we extend this to many moment-matching distributions concentrated at different locations. \Cref{app:Moment-lowerbound-search-app} then turns this family into a search lower bound. Finally, \Cref{app:moment-positive-app} complements these negative results with structural conditions under which moments are informative, namely, bounded point masses and small support.

\subsection{Two Distributions with Separated Peaks}
\label{app:two-distributions-app}

We start with a negative result: the next proposition shows that, even with the first $k$ moments fixed, two distributions can have most of their mass concentrated at two different positions, at distance of order $n/k$ from each other. Later, we will use this construction for an actual lower bound for our search problem.

\twoDistros*

\begin{figure}[t]
  \centering
  \includegraphics[width=\textwidth]{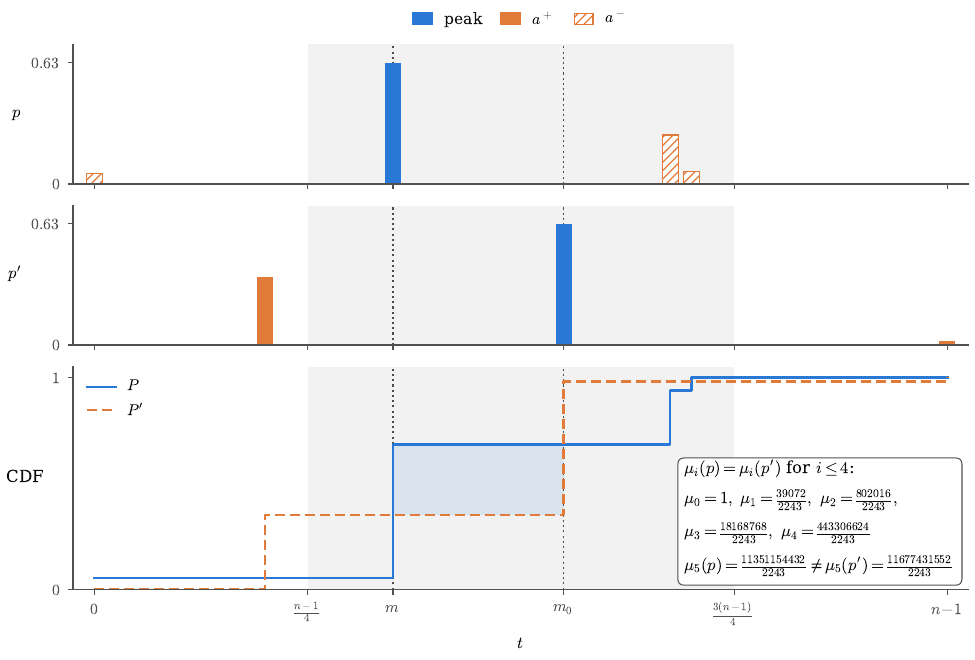}
  \caption{The construction of \Cref{prop:2Distros} for $n=41$, $k=4$, $m=14$ and $m_0=22$: the distributions $p=(\delta_m+a^-)/Z$ and $p'=(\delta_{m_0}+a^+)/Z$ have the same first $k$ moments, where $a$ is an optimal signed correction and $Z=1+\frac12\lVert a\rVert_1$. Bottom: their CDFs, which differ by $0.33$ on $[m,m_0)$; the box lists the common moments $\mu_0,\ldots,\mu_4$ and shows that $\mu_5$ already differs. Shading marks $[\frac{n-1}4,\frac{3(n-1)}4]$.}
  \label{2Distros-app}
\end{figure}

See \Cref{2Distros-app} for an example of this proposition. The proposition implies that matching only the first $k$ moments does not determine where most of the mass lies. Indeed, we can select positions $m,m_0$ in the middle section of $[n]_0$ with $|m-m_0|=\Theta(n/k)$, while keeping $\gamma\le 1/4$ and ensuring the size of the peaks is at least $3/4$. We discuss further consequences of the proposition at the end of \Cref{app:moment-positive-app}.

The proof of the proposition proceeds in four steps. We first rescale the ground set to a uniform grid in $[-1,1]$. We then reduce the construction to finding a signed correction vector with small $\ell_1$-norm. This smallest norm is the value of a linear program, whose dual asks for the largest increment $q(\tau_m)-q(\tau_{m_0})$ of a polynomial $q$ of degree $k$ that is bounded by $1$ on the grid. Finally, we bound this increment using Rakhmanov's theorem~\cite{rakhmanov2007} and Bernstein's inequality~\cite[Chapter~2]{rivlin1990}.%

\paragraph{Step 1: Rescaling the ground set}

Let $x(t):=\frac{2t-(n-1)}{n-1}$ be the affine map sending $[0,n-1]$ onto $[-1,1]$, and consider the corresponding scaled grid given by
\[
\tau_t:=x(t)=\frac{2t}{n-1}-1 \quad\text{for } t\in[n]_0,
\qquad
\mathcal{T}:=\{\tau_0,\ldots,\tau_{n-1}\}=\Bigl\{-1,-1+\tfrac2{n-1},\ldots,1-\tfrac2{n-1},1\Bigr\},
\]
whose consecutive points are at distance $2/(n-1)$. Note that $|\tau_m-\tau_{m_0}|=\frac2{n-1}|m-m_0|$ for all $m,m_0\in[n]_0$, and that $m\in[\frac {n-1}4,\frac{3(n-1)}4]$ if and only if $\tau_m\in[-\frac12,\frac12]$.

\begin{restatable}{lemma}{rescaling}\label{lm:rescaling}
For every $a\in\mathbb{R}^{n}$, the following are equivalent:
\begin{enumerate}[label=(\roman*)]
  \item $\sum_{t=0}^{n-1} a_t\, t^i=0$ for all $i=0,\ldots,k$;
  \item $\sum_{t=0}^{n-1} a_t\, \tau_t^i=0$ for all $i=0,\ldots,k$.
\end{enumerate}
In particular, two distributions $p,p'$ on $[n]_0$ have the same first $k$ moments if and only if $\sum_{t=0}^{n-1}p_t\tau_t^i=\sum_{t=0}^{n-1}p'_t\tau_t^i$ for all $i=0,\ldots,k$.
\end{restatable}
\begin{proof}
By linearity, (i) holds if and only if $\sum_{t=0}^{n-1} a_tf(t)=0$ for every polynomial $f$ of degree at most $k$, and (ii) holds if and only if $\sum_{t=0}^{n-1} a_t\,g(\tau_t)=0$ for every polynomial $g$ of degree at most $k$. Since $t=\frac {n-1}2(\tau_t+1)$, the substitution $g(y)=f\big(\frac {n-1}2(y+1)\big)$ is a bijection between polynomials of degree at most $k$, with inverse $f(t)=g(x(t))$, and it satisfies $f(t)=g(\tau_t)$ for all $t\in[n]_0$. Hence (i) and (ii) are equivalent. The last statement is the equivalence applied to $a=p-p'$.
\end{proof}

\paragraph{Step 2: Reduction to a signed correction}

For $v\in[n]_0$, let $\delta_v\in\mathbb{R}^n$ denote the unit point mass at $v$, i.e., $(\delta_v)_t=1$ if $t=v$ and $(\delta_v)_t=0$ otherwise. Given the peak positions $m,m_0\in[n]_0$, consider the linear system
\begin{equation}\label{eq:Vandermondsystem-app}
\sum_{t=0}^{n-1} a_t \tau_t^i = \tau_m^i-\tau_{m_0}^i \qquad \text{for all }i=0,\ldots,k,
\end{equation}
in the unknown $a\in\mathbb{R}^n$. Observe that $a$ solves~\Cref{eq:Vandermondsystem-app} if and only if the vector $a+\delta_{m_0}-\delta_m$ satisfies condition (ii) of \Cref{lm:rescaling}: %
Thus, adding $a$ to $\delta_{m_0}$ produces a signed vector with the same first $k$ moments as $\delta_m$.

\signedVectorReduction*

\begin{proof}
Let $a^+$ and $a^-$ be the positive and negative parts of $a$, that is,
\[
a_t^+=\max\{0,a_t\},\qquad a_t^-=\max\{0,-a_t\}\qquad\text{for } t\in[n]_0,
\]
so that $a=a^+-a^-$. Since the constraint for $i=0$ in~\Cref{eq:Vandermondsystem-app} gives $\sum_t a_t=0$, we have
\[
\sum_{t=0}^{n-1} a_t^+=\sum_{t=0}^{n-1} a_t^-=\frac12\lVert a\rVert_1.
\]
Define
\[
p:=\frac{\delta_m+a^-}{1+\frac12\lVert a\rVert_1}
\qquad\text{and}\qquad
p':=\frac{\delta_{m_0}+a^+}{1+\frac12\lVert a\rVert_1}.
\]
Both vectors are nonnegative and, by our previous argument, their entries sum to $1$; hence $p$ and $p'$ are probability distributions on $[n]_0$. Moreover, $p-p'=\bigl(\delta_m-\delta_{m_0}-a\bigr)/(1+\frac12\lVert a\rVert_1)$, and $a$ solves~\Cref{eq:Vandermondsystem-app} precisely when $\sum_t (\delta_m-\delta_{m_0}-a)_t\tau_t^i=0$ for all $i=0,\ldots,k$. By \Cref{lm:rescaling}, $p$ and $p'$ have the same first $k$ moments. Finally,
\[
p_m,\,p'_{m_0}\ge \frac{1}{1+\frac12\lVert a\rVert_1}
\ge 1-\frac12\lVert a\rVert_1.
\]
\end{proof}
\Cref{lm:signedVectorReduction} shows that, for given peak positions $m,m_0$, the question is how small the $\ell_1$-norm of a solution of~\Cref{eq:Vandermondsystem-app} can be. We therefore define
\begin{equation}\label{eq:gammastar-app}
\gamma^*(m,m_0):=\inf\bigl\{\lVert a\rVert_1 \,:\, a\in\mathbb{R}^{n} \text{ satisfies~\Cref{eq:Vandermondsystem-app}}\bigr\}.
\end{equation}
By \Cref{lm:signedVectorReduction}, every solution $a$ of~\Cref{eq:Vandermondsystem-app} yields distributions as in \Cref{prop:2Distros} with $\gamma=\frac12\lVert a\rVert_1$. It thus remains to bound $\gamma^*(m,m_0)$, which we do in the next two steps. %

\paragraph{Step 3: The dual of the correction problem}

Writing $a=a^+-a^-$ with $a^+,a^-\ge0$ as in the previous proof, the value $\gamma^*(m,m_0)$ is the optimum of the linear program [P] below, which we call the \emph{correction LP}:
\begin{align*}
        \text{\rm [P]}\qquad
        \min \quad & \sum_{t=0}^{n-1} (a_t^++a_t^-) \\
        \text{s.t.}\quad
        & \sum_{t=0}^{n-1} (a^+_t-a^-_t)\,\tau_t^i=\tau_m^i-\tau_{m_0}^i \qquad \text{for all } i=0,\ldots,k,\\
        & a_t^+,a_t^-\ge 0 \qquad \text{for all } t=0,\ldots,n-1.
\end{align*}
 The dual of [P] has a simple interpretation in terms of polynomials. Its dual is
\begin{align*}
        \text{\rm [D]}\qquad
        \max \quad & \sum_{i=0}^k \lambda_i \bigl(\tau_m^i-\tau_{m_0}^i\bigr)\\
        \text{s.t.}\quad
        & -1\le \sum_{i=0}^k \lambda_i \tau_t^i \le 1 \qquad \text{for all } t=0,\ldots,n-1,
\end{align*}
and, identifying a dual solution $(\lambda_0,\ldots,\lambda_k)$ with the polynomial $q(x)=\sum_{i=0}^k\lambda_i x^i$ of degree at most $k$, the constraints say that $|q(\tau_t)|\le1$ for all $t\in[n]_0$ and the objective is $q(\tau_m)-q(\tau_{m_0})$. We call such a $q$ a \emph{feasible polynomial}. The correction LP is feasible (e.g., $a^+=\delta_m$, $a^-=\delta_{m_0}$) and its objective is bounded below by $0$, so its optimum is attained, and by strong linear programming duality it coincides with the optimum of [D]. Together with the observation above that [P] computes $\gamma^*(m,m_0)$, this gives the following lemma.

\dualCorrection*

With this lemma, we get that to upper bound $\gamma^*(m,m_0)$, it suffices to upper bound $q(\tau_m)-q(\tau_{m_0})$ for every feasible polynomial. %

\paragraph{Step 4: Bounding the increments of feasible polynomials}

In order to bound $q(\tau_m)-q(\tau_{m_0})$, we bound the derivative of $q$ and use the mean value theorem. We use two classical inequalities for real polynomials $q$ with $\deg q\le k$; see, e.g.,~\cite[Chapter~2]{rivlin1990}. Writing $\lVert q\rVert:=\max_{x\in[-1,1]}|q(x)|$,
\begin{align}
  \text{(Markov)}\qquad &\max_{x\in[-1,1]}|q'(x)|\;\le\;k^{2}\,\lVert q\rVert,
  \label{eq:markov-app}\\
  \text{(Bernstein)}\qquad &|q'(x)|\;\le\;\frac{k}{\sqrt{1-x^{2}}}\,\lVert q\rVert
  \qquad \text{for all } x\in(-1,1). \label{eq:bernstein-app}
\end{align}
Applying~\Cref{eq:bernstein-app} to $x\mapsto q(\rho x)$ shows that, for every $\rho>0$,
\begin{equation}\label{eq:bernsteinScaled-app}
  |q(x)|\le B \text{ for all } x\in[-\rho,\rho]
  \quad\Longrightarrow\quad
  |q'(x)|\le\frac{k\,B}{\sqrt{\rho^2-x^2}}\quad\text{for all } x\in(-\rho,\rho).
\end{equation}

The constraint in \Cref{lem:dualCorrection} bounds $q$ on the grid $\mathcal{T}$ only, whereas~\Cref{eq:markov-app} and~\Cref{eq:bernstein-app} require a bound on an interval. As a warm-up, we show that when $k$ is small compared to $\sqrt n$, the transfer follows from Markov's inequality. We will not use this lemma later; it serves to illustrate the issue, which is handled in full generality by \Cref{thm:rakhmanov} below.

\begin{restatable}[From the grid to the interval]{lemma}{gridToInterval}\label{lem:grid}
Let $q$ be a polynomial with $\deg q\le k$ and $|q(\tau_t)|\le1$ for all $t\in[n]_0$. If $n-1\ge 2k^{2}$, then $\lVert q\rVert\le 2$.
\end{restatable}

\begin{proof}
Let $\lVert q\rVert=|q(x^\ast)|$ with $x^\ast\in[-1,1]$, and let $\tau\in\mathcal T$ be a point with $|x^\ast-\tau|\le 1/(n-1)$. By~\Cref{eq:markov-app} and the mean value theorem,
\[
  \lVert q\rVert\;\le\;|q(\tau)|+|x^\ast-\tau|\max_{x\in[-1,1]}|q'(x)|
  \;\le\;1+\frac{k^{2}}{n-1}\,\lVert q\rVert ,
\]
so $\lVert q\rVert\le(1-k^{2}/(n-1))^{-1}\le 2$.
\end{proof}

\Cref{lem:grid}, however, requires $n-1\ge2k^2$. Relying on it alone, our construction would only yield the distributions $p,p'$ of \Cref{prop:2Distros} for $k=O(\sqrt n)$, and we could not rule out a constant-factor approximate search strategy that knows $n^{1/2+\varepsilon}$ moments. However, the hypothesis $n-1\ge 2k^2$ can be dropped if one only asks for a bound on the central part of the interval, away from the endpoints $\pm1$: there, polynomials bounded on the grid remain bounded by an absolute constant on an interval that depends only on the ratio $k/n$, as shown by the following landmark theorem of Rakhmanov.%

\begin{restatable}[{Rakhmanov~\cite[Theorem~1]{rakhmanov2007}}]{theorem}{rakhmanovTheorem}\label{thm:rakhmanov}
Let $E_N=\{\zeta_j=-1+(2j-1)/N:\ j=1,\ldots,N\}$ be the set of $N$ equidistant points in $(-1,1)$, let $\kappa<N$, and let $Q$ be a polynomial of degree at most $\kappa$ with $\max_{\zeta\in E_N}|Q(\zeta)|\le1$. Let $r:=\sqrt{1-\kappa^{2}/N^{2}}$. Then, there exists an absolute constant $C_0$ such that
\[
  |Q(x)|\;\le\;C_0\log\frac{\pi}{\arctan\bigl(\frac N\kappa\sqrt{r^{2}-x^{2}}\bigr)}
  \qquad\text{for all } x\in(-r,r).
\]
\end{restatable}

Our grid $\mathcal T$ is not exactly $E_n$, as it contains the endpoints $\pm1$, but it is a dilation of it: for $j=1,\ldots,n$, $\frac{n}{n-1} \zeta_j=\tau_{j-1}.$ The next lemma gives an interpretation of Rakhmanov's result for $\mathcal T$ in a form that is useful for us.

\gridToCentralInterval*

\begin{proof}
The case $k=0$ is immediate. Otherwise, set $Q(y):=q(\frac{n}{n-1}y)$.
Since $Q(\zeta_j)=q(\tau_{j-1})$, \Cref{thm:rakhmanov} applies with $N=n$ and $\kappa=k$.

Fix $x\in[-3/4,3/4]$ and set $y=(n-1)x/n$, so that $Q(y)=q(x)$ and $|y|\le3/4$.
Since $k/n<1/2$, we have $r^2=1-k^2/n^2>3/4$, and hence
\[
r^2-y^2>\frac34-\frac9{16}=\frac3{16},
\qquad
\frac nk\sqrt{r^2-y^2}>\frac{\sqrt3}{2}.
\]
Thus \Cref{thm:rakhmanov} and monotonicity of $\arctan$ give
\[
|q(x)|=|Q(y)| \le C_0\log\frac{\pi}{\arctan(\sqrt3/2)} =:C_1. \qedhere 
\]
\end{proof}

We can now bound the value of the correction problem.

\cheapCorrection*

\begin{proof}
Let $q$ be an optimal solution of the maximization problem in \Cref{lem:dualCorrection}, so that $\gamma^*(m,m_0)=q(\tau_m)-q(\tau_{m_0})$. Recall that $\tau_m,\tau_{m_0}\in[-\frac12,\frac12]$.

By \Cref{lem:gridCentral}, $|q|\le C_1$ on $[-\frac34,\frac34]$. Hence~\Cref{eq:bernsteinScaled-app} with $\rho=\frac34$ gives, for all $x\in[-\frac12,\frac12]$,
\[
|q'(x)|\le\frac{k\,C_1}{\sqrt{\frac9{16}-\frac14}}=\frac{4C_1}{\sqrt5}\,k.
\]
By the mean value theorem,
\[
\gamma^*(m,m_0)=q(\tau_m)-q(\tau_{m_0})\le\frac{4C_1}{\sqrt5}\,k\,|\tau_m-\tau_{m_0}|=\frac{8C_1}{\sqrt5}\,\frac{k\,|m-m_0|}{n-1}.
\]

\end{proof}

\begin{proof}[Proof of \Cref{prop:2Distros}]
Let $a$ be a minimizer of~\Cref{eq:gammastar-app}, so that $\lVert a\rVert_1=\gamma^*(m,m_0)$. Set $C:=4C_1/\sqrt5$. By \Cref{lem:cheapCorrection}, $\lVert a\rVert_1\le 2C\,\frac{k}{n-1}|m-m_0|$. \Cref{lm:signedVectorReduction} then yields two distributions $p,p'$ on $[n]_0$ with matching first $k$ moments and
\[
p_m,\,p'_{m_0}\ge 1-\tfrac12\lVert a\rVert_1\ge1-C\,\frac{k}{n-1}|m-m_0|.
\]
This proves the proposition.
\end{proof}

\subsection{Several Distributions with the Same Moments}
\label{app:many-distributions-app}

\begin{figure}[t]
  \centering
  \includegraphics[width=\textwidth]{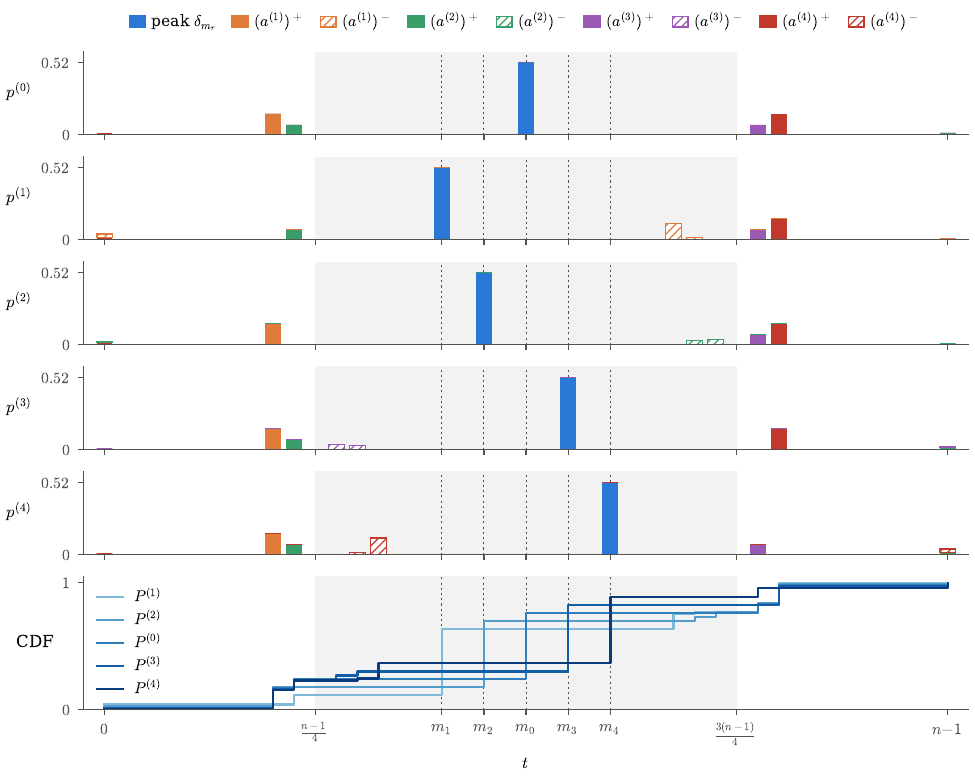}
  \caption{The construction of \Cref{prop:manyDistrosSameMoments} for $n=41$, $k=4$ and peaks $(m_0,\ldots,m_4)=(20,16,18,22,24)$; the five distributions have the same first $k$ moments. Each $p^{(r)}$ contains $(a^{(s)})^+$ for all $s\neq r$ and, for $r\ge1$, contains $(a^{(r)})^-$ instead of $(a^{(r)})^+$. Bottom: their CDFs. Shading marks $[\frac{n-1}4,\frac{3(n-1)}4]$.}
  \label{manyDistros-app}
\end{figure}

We now generalize the construction from two distributions to several distributions with the same first $k$ moments. To get distributions with several peaks at $m_0,m_1,\ldots,m_R$, we construct vectors $a^{(s)}\in\mathbb{R}^{n}$ optimizing~\Cref{eq:gammastar-app} for the pair $(m_s,m_0)$, and then combine them appropriately; see \Cref{manyDistros-app}.
 The cost is to obtain smaller peaks.

\manyDistrosSameMoments*

\begin{proof}
We use the signed-correction construction from the proof of \Cref{prop:2Distros}. For $s=1,\ldots,R$, let $a^{(s)}\in\mathbb{R}^{n}$ be a minimizer of~\Cref{eq:gammastar-app} for the pair $(m_s,m_0)$. By \Cref{lem:cheapCorrection},
\[
\lVert a^{(s)}\rVert_1\le 2C\,\frac k{n-1}|m_s-m_0| \qquad\text{for all } s=1,\ldots,R.
\]
Write $a^{(s)}=(a^{(s)})^+-(a^{(s)})^-$, where $(a^{(s)})^{\pm}\ge0$ are the positive and negative parts of $a^{(s)}$. Since the constraint for $i=0$ in~\Cref{eq:Vandermondsystem-app} gives $\sum_t a^{(s)}_t=0$, we have $\lVert(a^{(s)})^+\rVert_1 = \lVert(a^{(s)})^-\rVert_1 = \frac12\lVert a^{(s)}\rVert_1.$

Define $B:=\sum_{s=1}^R (a^{(s)})^+$ and $Z:=1+\lVert B\rVert_1$, and let
\[
p^{(0)} := \frac{\delta_{m_0}+B}{Z}
\qquad\text{and}\qquad
p^{(r)} := \frac{\delta_{m_r} + (a^{(r)})^- + \sum_{s\neq r}(a^{(s)})^+ }{Z}
\quad\text{for } r=1,\ldots,R.
\]
All these vectors are nonnegative. Moreover, simple algebra shows that the numerator of $p^{(0)}$ and $p^{(r)}$, for all $r\ge 1$,  has total mass $Z$.
Thus $p^{(0)},\ldots,p^{(R)}$ are probability distributions on $[n]_0$.

We now check the moments. For $r\ge1$, the numerators of $p^{(0)}$ and $p^{(r)}$ differ by
\[
(\delta_{m_0}+B)-\Bigl(\delta_{m_r} + (a^{(r)})^- + \sum_{s\neq r}(a^{(s)})^+\Bigr)
= \delta_{m_0}-\delta_{m_r}+(a^{(r)})^+-(a^{(r)})^-
= a^{(r)}+\delta_{m_0}-\delta_{m_r},
\]
which satisfies condition (ii) of \Cref{lm:rescaling} because $a^{(r)}$ solves~\Cref{eq:Vandermondsystem-app} for the pair $(m_r,m_0)$. By \Cref{lm:rescaling}, $p^{(0)}$ and $p^{(r)}$ have the same first $k$ moments. Hence all distributions $p^{(0)},p^{(1)},\ldots,p^{(R)}$ have the same first $k$ moments.

Finally, the claim follows since $p^{(r)}_{m_r} \ge 1/Z$ for all $r=0,\ldots,R$, and
\[
\frac1Z \ge 1-\lVert B\rVert_1 = 1-\frac12\sum_{s=1}^R \lVert a^{(s)}\rVert_1
\ge 1- C\,\frac k{n-1}\sum_{s=1}^R |m_s-m_0|.
\]
\end{proof}

\subsection{A Search Lower Bound from Moment Matching}\label{app:Moment-lowerbound-search-app}
The following theorem establishes that even $n^{1-\delta}$ exact moments leave a worst-case
approximation ratio of $\Omega(\log n)$, which is meaningless as this ratio can be obtained with no information via binary search. This lies in stark contrast to \Cref{cor:searchFromDyadicJL}, which establishes that
$\widetilde O(n^\varepsilon)$ appropriately chosen linear measurements suffice for an $O(1/\varepsilon)$-approximation.

\momentSearchSeparation*
The lower bound also holds for randomized search strategies; see~\Cref{sec:randomized-lower-bounds}.

\begin{proof}
Let $C$ be the constant of \Cref{prop:manyDistrosSameMoments}, let $C':=8C$, and set
\[
R:=\left\lfloor \sqrt{\frac{n-1}{4C\,k\log n}}\right\rfloor.
\]
Since $n-1\ge n/2$, the assumption on $k$ gives $R\ge1$; moreover, $R\le\sqrt n$.
Let $m_0:=\lfloor (n-1)/2\rfloor$ and $m_r:=m_0+r$ for $r=1,\ldots,R$.
For all sufficiently large $n$, these points lie in $[\frac {n-1}4,\frac{3(n-1)}4]$, and $k\le (n-1)/2$. Hence, \Cref{prop:manyDistrosSameMoments} gives distributions
$p^{(0)},\ldots,p^{(R)}$ with the same first $k$ moments and
\[
p^{(r)}_{m_r}\ge 1-\gamma,
\qquad
\gamma
\le C\frac{k}{n-1}\sum_{s=1}^R s
\le C\frac{kR^2}{n-1}
\le \frac{1}{4\log n}.
\]

Since all these distributions have the same first $k$ moments, they lie in a
common uncertainty set $U\in\mathcal U_{\sigma_k}$.
Fix any deterministic search strategy $T$ for $U$, and let
$d_r:=q(T,m_r)$ be the number of queries $T$ uses to identify $m_r$. Represent $T$
as a binary search tree in which a target $v$ is found at the node querying $v$,
at cost one plus the depth of that node; a target that is the only remaining
candidate is found at a leaf without a further query, at cost equal to its depth.
Attaching to each node that queries $v$ an extra leaf for the outcome $t=v$ yields
a ternary tree in which every target ends at its own leaf, at depth equal to its
number of queries. Kraft's inequality~\cite[Theorem~5.2.1]{cover_elements_2006}
for this tree gives $\sum_{r=0}^R 3^{-d_r}\le1$, and by convexity of
$x\mapsto3^{-x}$, Jensen's inequality yields
\[
\frac{1}{R+1}\sum_{r=0}^R d_r\ge \log_3(R+1).
\]
Thus there is an index $r\in\{0,\ldots,R\}$ for which
$d_r\ge\log_3(R+1)$. Hence the expected cost of $T$ on
$p^{(r)}$ is at least $(1-\gamma)\log_3(R+1)=\Omega(\log (R+1)).$

On the other hand, a strategy that knows $p^{(r)}$ can first query the vertex
$m_r$. This identifies $m_r$ with a single query; otherwise, the strategy finishes
by ordinary binary search on the side containing the target, which uses at most
$\lceil\log_2 n\rceil$ further queries. Since the total mass outside $m_r$ is at
most $\gamma$, this gives
\[
\opt(p^{(r)})
\le 1+\gamma\,\lceil\log_2 n\rceil
=O(1).
\]
It follows that every strategy for $U$ has an approximation ratio $\Omega(\log (R+1))$ on some distribution in $U$, and hence $\alpha(\mathcal U_{\sigma_k})=\Omega(\log(R+1))$. Finally, since $R\ge1$ and $R+1\ge\sqrt{(n-1)/(4Ck\log n)}\ge\sqrt{n/(8Ck\log n)}$, we have that $\log(R+1)=\Omega\Bigl(\log\frac{n}{k\log n}\Bigr).$
If $k\le n/\log^2n$, then $\log n\le\sqrt{n/k}$, so $\frac{n}{k\log n}\ge\sqrt{n/k}$ and the bound is $\Omega(\log(n/k))$. If $k\le n^{1-\delta}$, then $k\le n/\log^2 n$ for $n$ large enough and $\log(n/k)\ge\delta\log n$.
\end{proof}

\subsection{Approximate and Exact Recovery from Moments}
\label{app:moment-positive-app}

We also give self-contained proofs of recovery guarantees under bounded-atom and small-support assumptions.
If the observed moments admit a nominal distribution with small
atoms, we bound the CDF distance between that distribution and
every other distribution with the same moments.
If they admit a nominal distribution with small support,
sufficiently many moments determine the distribution uniquely.
Both conclusions concern the full uncertainty set of distributions
consistent with the observed moments.

Both arguments work under the same principle: we consider an LP to compute the maximum/minimum values attainable by the CDF, and then create a dual certificate using polynomial interpolation techniques. It is worth mentioning that the LP we consider is the finite dimensional version of the semi-infinite linear program considered by Bertsimas and Popescu~\cite{bertsimasPopescu2005} and Bertsimas and Sethuraman~\cite{bertsimas_moment_2000}, but also considered implicitly in the classical literature of moments; see, e.g., the monograph by Karlin and Studden~\cite{karlin1966tchebycheff}. 

To introduce the considered linear program, fix the moments $\bar\mu_0,\ldots,\bar\mu_k$ of the (unknown) true distribution, and let $\pi$ be a nominal distribution with these moments, that is, $\bar\mu_i=\sum_{t=0}^{n-1} t^i \pi_t$ for $i=0,\ldots,k$; such a $\pi$ can be computed from the moments by solving a linear feasibility problem. For $M\in\{0,\ldots,n-1\}$, the largest possible CDF value at $M$, among all distributions with these moments, is given by
\begin{align*}
        \text{\rm [P$'$]}\qquad
        \max \quad & \sum_{t=0}^M p_t \\
        \text{s.t.}\quad
        & \sum_{t=0}^{n-1} t^i p_t=\bar\mu_i \qquad \text{for all } i=0,\ldots,k,\\
        & p_t\ge 0 \qquad \text{for all } t=0,\ldots,n-1.
\end{align*}
The corresponding dual is
\begin{align*}
        \text{\rm [D$'$]}\qquad
        \min \quad & \sum_{i=0}^k \lambda_i \bar\mu_i\\
        \text{s.t.}\quad
        & \sum_{i=0}^k \lambda_i t^i \ge 1 \qquad \text{for } t=0,\ldots,M,\\
        & \sum_{i=0}^k \lambda_i t^i \ge 0 \qquad \text{for } t=M+1,\ldots,n-1.
\end{align*}
Hence, as for [D] in \Cref{app:two-distributions-app}, a dual solution $(\lambda_0,\ldots,\lambda_k)$ can be identified with a polynomial
$q(x)=\sum_{i=0}^k\lambda_i x^i$ of degree at most $k$. Thus, a feasible dual solution is a polynomial satisfying
\[
q(t)\ge 1 \quad \text{for } t\in \{0,1,\ldots,M\},
\qquad \text{and }
q(t)\ge 0 \quad \text{for } t\in \{M+1,\ldots,n-1\}.
\]
Moreover, we can rewrite the dual objective as
\[
\sum_{i=0}^k \lambda_i \bar\mu_i
= \sum_{i=0}^k \lambda_i \sum_{t=0}^{n-1} t^i \pi_t
= \sum_{t=0}^{n-1} \pi_t \sum_{i=0}^k \lambda_i t^i
= \sum_{t=0}^{n-1} \pi_t\, q(t).
\]
Hence, the dual seeks a polynomial $q$ of degree at most $k$ such that $q(t)\ge 0$ for all $t\in\{0,\ldots,n-1\}$, $q(t)\ge 1$ for all $t\in\{0,\ldots,M\}$, and minimizes $\sum_{t=0}^{n-1} \pi_t\, q(t)$.

As a warm-up, consider $k=n-1$. By Lagrange interpolation there is a polynomial $q$ of degree at most $n-1$ with $q(t)=1$ for $t\in\{0,\ldots,M\}$ and $q(t)=0$ for $t\in\{M+1,\ldots,n-1\}$. It is dual feasible with objective value $\sum_{t=0}^M \pi_t$, the mass that $\pi$ places on $\{0,\ldots,M\}$. By weak duality, no distribution with the same moments places more mass on $\{0,\ldots,M\}$. Applying the same argument to the minimization version of [P$'$] shows that none places less. Hence the first $n-1$ moments determine the CDF, and thus the distribution itself; this also follows from the invertibility of the Vandermonde matrix. When $k$ is much smaller than $n$, exact interpolation is no longer possible. \Cref{lem:cert} shows that we can still construct a feasible dual polynomial that smoothly approximates $\indic_{\{t\le M\}}$, and whose objective value is (in some sense) approximately $\sum_{t=0}^M \pi_t$.

\begin{figure}[t]
  \centering
  \includegraphics[width=\textwidth]{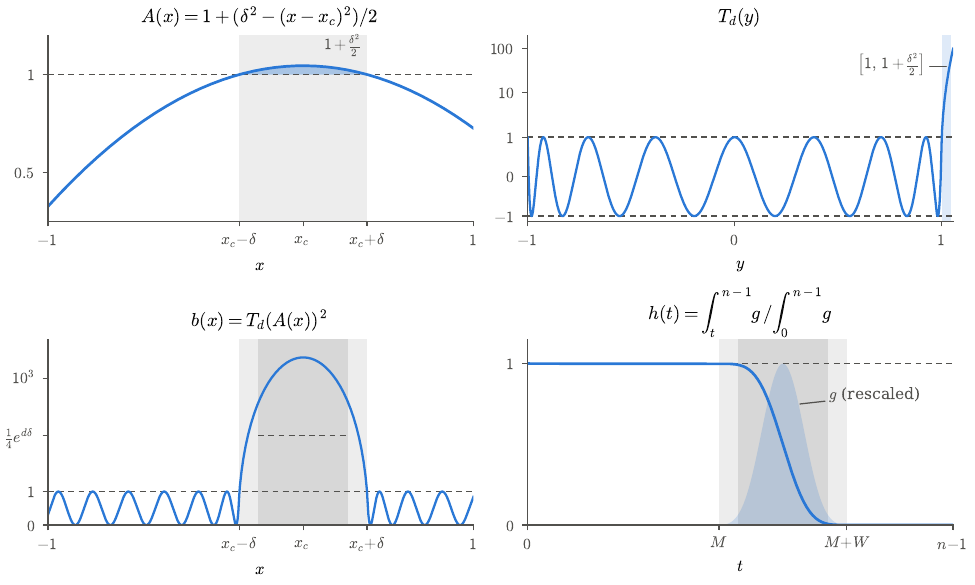}
  \caption{Illustrations of the constructions in \Cref{lem:bump,lem:cert}. Shading marks $|x-x_c|\le\delta$ (light) and $|x-x_c|\le\delta/\sqrt2$ (dark). The plots of $T_d$ and $b$ are logarithmic above $1$.}
  \label{bump-app}
\end{figure}

The construction has two steps. We first build, from a Chebyshev polynomial, a polynomial bump on $[-1,1]$ that is concentrated on a short interval (\Cref{lem:bump}; see \Cref{bump-app} for an example). We then shift and scale the bump to the window $[M,M+W]$, integrate it to obtain a smoothed step (see the bottom-right panel of \Cref{bump-app}), and rescale the result to obtain the certificate (\Cref{lem:cert}).

We recall the few facts about Chebyshev polynomials that we use; see~\cite[Chapter~3]{trefethen2019atap} or \cite[Chapters 3 and 4]{carothers_short_1998} for modern references. For every integer $d\ge0$, $\cos(d\theta)$ is a polynomial of degree $d$ in $\cos\theta$, and the \emph{Chebyshev polynomial of the first kind} $T_d$ is this polynomial, that is, $T_d(\cos\theta)=\cos(d\theta)$ for all $\theta\in\mathbb{R}$; see the top-right panel of \Cref{bump-app} for an example. %

Two consequences are used in the next lemma. First, writing $x=\cos\theta$ shows that $|T_d(x)|\le1$ for $x\in[-1,1]$. Observe, however, that the relation $T_d(\cos(\theta))=\cos(d\theta)$ only tells us something about $T_d(x)$ for $x\in [-1,1]$. To study $T_d$ outside this range, one can consider $y=\cos(i\theta)=\cosh(\theta)$, which also satisfies the relation $T_d(y)=T_d(\cos(i\theta)) = \cos(id\theta) = \cosh(d\theta)$, implying that $T_d(y)=\cosh(d\operatorname{arccosh}y)$ for $y\ge1$. Hence, in this range $T_d(y)$ is increasing and equal to
\[T_d(y)=\frac{e^{d\operatorname{arccosh}y} + e^{-d\operatorname{arccosh}y}}{2}\ge \frac{e^{d\operatorname{arccosh}y}}{2}.\] As $\operatorname{arccosh}(1+\varepsilon)\approx\sqrt{2\varepsilon}$ for small $\varepsilon>0$, the polynomial $T_d$ is bounded by $1$ on $[-1,1]$ but already of order $e^{d\sqrt{2\varepsilon}}$ at $1+\varepsilon$. \Cref{lem:bump} exploits this contrast by composing $T_d$ with a quadratic map to localize a bump arbitrarily.%

\begin{restatable}[Chebyshev bump]{lemma}{chebyshevBump}\label{lem:bump}
Let $d\ge1$ be an integer, $x_c\in[-1,1]$ and $\delta\in(0,\tfrac12]$, and define
\[
  A(x)=1+\frac{\delta^{2}-(x-x_c)^{2}}{2},
  \qquad
  b(x)=T_d\big(A(x)\big)^{2},
\]
where $T_d$ is the Chebyshev polynomial of the first kind of degree $d$. Then $b$ is a polynomial of degree $4d$ such that
\begin{enumerate}[label=(\alph*)]
  \item $b(x)\ge0$ for all $x\in\mathbb{R}$;
  \item $b(x)\le 1$ for all $x\in[-1,1]$ with $|x-x_c|\ge\delta$;
  \item $b(x)\ge \tfrac14\,e^{d\delta}$ for all $x$ with $|x-x_c|\le\delta/\sqrt2$.
\end{enumerate}
\end{restatable}

\begin{proof}
It's easy to see that $b$ has degree $4d$. Property (a) holds because $b$ is a square.

For (b), let $x\in[-1,1]$ with $|x-x_c|\ge\delta$. Simple algebra shows that $|A(x)|\le1$, and since $|T_d|\le1$ on $[-1,1]$, we get $b(x)=T_d(A(x))^2\le1$.

For (c), let $|x-x_c|\le\delta/\sqrt2$, so that $A(x)\ge1+\delta^2/4$. For $y\ge1$ we have $T_d(y)=\cosh\big(d\operatorname{arccosh}y\big)$, which is increasing in $y$, and $\cosh u\ge\tfrac12e^{u}$. Furthermore, since $y+1\le4$ for $y\in[1,1+\delta^2/4]$,
\[
  \operatorname{arccosh}\Big(1+\frac{\delta^2}{4}\Big)
  =\int_1^{1+\delta^2/4}\frac{dy}{\sqrt{(y-1)(y+1)}}
  \ge\int_1^{1+\delta^2/4}\frac{dy}{2\sqrt{y-1}}
  =\frac{\delta}{2}.
\]
Hence
\[
  T_d\big(A(x)\big)\ge T_d\Big(1+\frac{\delta^2}{4}\Big)\ge\tfrac12e^{d\delta/2},
\]
and squaring yields (c).
\end{proof}

\smoothedStep*

\begin{proof}
We start by shifting the bump to the window. Set $d=\lfloor k/8\rfloor$, $\delta=W/(n-1)\in(0,\tfrac12]$ and $c=M+W/2$. Let $x(t)=(2t-(n-1))/(n-1)$ be the affine map sending $[0,n-1]$ onto $[-1,1]$, and let $x_c=x(c)$; since $0\le M\le c\le M+W\le n-1$, we have $x_c\in[-1,1]$. Let $b$ be the polynomial of \Cref{lem:bump} with parameters $d$, $x_c$, and $\delta$, and define $g(t)=b(x(t))$. Then $g$ is a polynomial of degree $4d$ in $t$, and $g\ge0$ by \Cref{lem:bump}(a). Since $|x(t)-x_c|=2|t-c|/(n-1)$, the condition $|x(t)-x_c|\ge\delta$ is equivalent to $|t-c|\ge W/2$, that is, to $t\le M$ or $t\ge M+W$. Hence \Cref{lem:bump}(b) gives
\begin{equation}\label{eq:outside-app}
  g(t)\le 1 \qquad\text{for } t\in[0,M]\cup[M+W,n-1].
\end{equation}
Similarly, $|x(t)-x_c|\le\delta/\sqrt2$ is equivalent to $|t-c|\le W/(2\sqrt2)$, an interval of length $W/\sqrt2$ contained in $[M,M+W]\subseteq[0,n-1]$. On this interval $g\ge\tfrac14e^{d\delta}$ by \Cref{lem:bump}(c), whence
\begin{equation}\label{eq:inside-app}
  \int_0^{n-1} g \;\ge\; \frac{W}{\sqrt2}\cdot\frac{e^{d\delta}}{4}
  \;\ge\;\frac{W}{6}\,e^{d\delta}.
\end{equation}
To obtain an approximate step function, we integrate and normalize $g$. Put
\[
  h(t)=\frac{\int_t^{n-1} g(s)\,ds}{\int_0^{n-1} g(s)\,ds},
  \qquad \deg h=4d+1 .
\]
Since $g\ge0$, $h$ is nonincreasing on $[0,n-1]$ with $h(0)=1$, $h(n-1)=0$ and $0\le h\le1$.
By \Cref{eq:outside-app} both $\int_0^M g$ and $\int_{M+W}^{n-1} g$ are at most $n-1$, so
with \Cref{eq:inside-app},
\begin{equation}\label{eq:tails-app}
  1-h(M)=\frac{\int_0^M g}{\int_0^{n-1} g}\le\frac{6(n-1)}{W}e^{-d\delta},
  \qquad
  h(M+W)=\frac{\int_{M+W}^{n-1} g}{\int_0^{n-1} g}\le\frac{6(n-1)}{W}e^{-d\delta}.
\end{equation}
Now $d\ge k/16$ (as $k\ge16$) and $\delta=W/(n-1)\ge W/n$, so
$d\delta\ge kW/(16n)\ge 6\ln n$ by the hypothesis on $W$. As $W\ge1$, both quantities in
\Cref{eq:tails-app} are at most $6(n-1)n^{-6}\le 6n^{-5}\le n^{-3}$, where the last step uses $n\ge3$.

\smallskip
Finally, we rescale $h$ so that it is at least $1$ on $[0,M]$. Define
\[
  q=\frac{h}{h(M)},\qquad \deg q=4d+1\le \tfrac k2+1\le k .
\]
(i) holds because $h\ge0$ on $[0,n-1]$. (ii) holds because $h$ is nonincreasing:
$t\le M$ gives $h(t)\ge h(M)$. For (iii), $h\le1$ and $h(M)\ge1-n^{-3}$ give
$q\le h(M)^{-1}\le(1-n^{-3})^{-1}\le1+n^{-2}$. For (iv), $t\ge M+W$ gives
$h(t)\le h(M+W)\le n^{-3}$, so $q(t)\le n^{-3}(1-n^{-3})^{-1}\le n^{-2}$.
\end{proof}

The next theorem quantifies how small atoms in a nominal
distribution limit the CDF ambiguity left by its first $k$ moments.
The bound applies to every distribution with the same moments,
without imposing the same atom bound on those distributions. The proof relies on the dual construction in \Cref{lem:cert} and weak duality.

\momentRecoveryBoundedAtoms*

\begin{proof}
Let $W:=\lceil 96\,(n/k)\ln n\rceil$. We may assume that $W\le(n-1)/2$, as otherwise $96\,(n/k)\ln n=\Omega(n)$, and since $L\ge\|\pi\|_\infty\ge1/n$ we get $L\,(n/k)\ln n=\Omega(1)$, so the claim holds trivially.

Fix $M\in\{0,\ldots,n-1\}$. We first show that
\begin{equation}\label{eq:cdfUpper-app}
P(M)\le\Pi(M)+(W-1)L+n^{-2}.
\end{equation}
If $M+W\ge n$, this holds because $P(M)\le1=\Pi(M)+\sum_{t=M+1}^{n-1}\pi_t\le\Pi(M)+(n-1-M)L$ and $n-1-M\le W-1$. Otherwise $M+W\le n-1$, and \Cref{lem:cert} gives a polynomial $q$ of degree at most $k$ satisfying (i)--(iv). By (i) and (ii), $q$ is feasible for {\rm [D$'$]}, and by (iii) and (iv) its objective value is
\[
\sum_{t=0}^{n-1}\pi_t\,q(t)\le(1+n^{-2})\sum_{t=0}^{M+W-1}\pi_t+n^{-2}\sum_{t=M+W}^{n-1}\pi_t=\Pi(M+W-1)+n^{-2}.
\]
Since $p$ is feasible for {\rm [P$'$]}, weak duality gives $P(M)\le\Pi(M+W-1)+n^{-2}$. As $\pi$ places mass at most $L$ on each of the $W-1$ points $M+1,\ldots,M+W-1$, this implies \Cref{eq:cdfUpper-app}.

For the reverse inequality, consider the reflected distributions $\tilde\pi_t:=\pi_{n-1-t}$ and $\tilde p_t:=p_{n-1-t}$\footnote{Alternatively, one could consider the corresponding minimization problem of [P$'$] and reflect the dual construction.}. They also have the same first $k$ moments, since $(n-1-t)^i$ is a polynomial of degree $i$ in $t$, and $\|\tilde\pi\|_\infty\le L$. Applying \Cref{eq:cdfUpper-app} to $\tilde\pi,\tilde p$ at the point $n-2-M$ gives $1-P(M)\le1-\Pi(M)+(W-1)L+n^{-2}$ (for $M=n-1$ this is trivial, as both CDFs equal $1$). Hence
\[
d_K(\pi,p)\le(W-1)L+n^{-2}\le 96\,L\,\frac{n}{k}\ln n+n^{-2}\le 97\,L\,\frac{n}{k}\ln n,
\]
where the last step uses $n^{-2}\le \ln n/n\le L\,(n/k)\ln n$, as $L\ge1/n$ and $k\le n$.

\end{proof}

Some dependency on $L$ in the theorem is necessary: for some $k=\Omega(n)$, \Cref{prop:2Distros} yields two distributions with $k$ matching moments, a peak of mass $\Omega(1)$, and Kolmogorov distance $\Omega(1)$. Hence, no bound independent of $L$ can tend to $0$, even with linearly many moments. In contrast, for $k=\Theta(n)$, our bound tends to $0$ whenever $L=o(1/\log n)$.

\begin{remark}
\Cref{thm:momentRecoveryBoundedAtoms} gives a quantitative CDF-recovery bound. In the parameter regime where its bound guarantees $d_K(\pi,p)\le n^{-\eps}$, we also have $L=O(n^{-\eps})$. Every compatible distribution therefore satisfies $p_t\le\pi_t+2d_K(\pi,p)=O(n^{-\eps})$ for every $t$. Thus, $H(p)=\Omega(\eps\log n)$, and ordinary binary search already achieves an $O(1/\eps)$-approximation in this regime.
\end{remark}

Finally, we recall that if a nominal distribution with the observed moments has support of size $\ell$, then $k=2\ell$ moments determine the true distribution exactly (it coincides with the nominal one), and therefore we can optimize without error.

This is the known identifiability result of Wu and Yang~\cite{wuYang2020}; we give the proof in our LP-duality framework for completeness.

\momentRecoverySmallSupport*

\begin{proof}
It suffices to show that the CDFs of $\pi$ and $p$ agree at every point. Fix $M\in\{0,\ldots,n-1\}$, and let $S:=\operatorname{supp}(\pi)\subseteq[n]_0$. Let $h_1$ be the unique polynomial of degree at most $\ell-1$ with
\[
h_1(t)=\indic_{\{t\le M\}}\qquad\text{for all }t\in S,
\]
and let $h_2(t):=\prod_{s\in S}(t-s).$ Then $h_2$ has degree $\ell$, vanishes on $S$, and is nonzero on $[n]_0\setminus S$. For $B>0$ sufficiently large, define
\[
q:=h_1+ B\,h_2^2.
\]
This polynomial has degree at most $2\ell\le k$ and agrees with $\indic_{\{t\le M\}}$ on $S$. Since $[n]_0$ is finite and $h_2(t)^2>0$ for $t\in[n]_0\setminus S$, as $B$ is sufficiently large, we have that $q(t)\ge\indic_{\{t\le M\}}$ for all $t \in  [n]_0$.
Thus $q$ is feasible for {\rm [D$'$]} and has objective value
\[
\sum_{t=0}^{n-1}\pi_t\,q(t)=\sum_{t\in S}\pi_t\,\indic_{\{t\le M\}}=\Pi(M).
\]
By weak duality, every distribution $p$ with the same first $k$ moments as $\pi$ satisfies $P(M)\le\Pi(M)$. By looking at the reflected distribution as in the proof of \Cref{thm:momentRecoveryBoundedAtoms}, we also have the reverse inequality. Hence $P=\Pi$.
\end{proof}

The preceding results show that moments can be informative under strong structural assumptions: either some nominal distribution with the given moments has sufficiently small atoms, or one has very small support. As shown in \Cref{app:two-distributions-app,app:Moment-lowerbound-search-app}, without such assumptions the first $k$ moments may be compatible with distributions whose mass is concentrated at very different locations, which makes it impossible to search with a close-to-optimal algorithm.

\Cref{thm:momentRecoveryBoundedAtoms} yields a tool to estimate the error of approximating a distribution based on $k$ of its moments. In order to guarantee a good approximation, however, we must have that some 
of the distributions $\pi$ that have the corresponding moments have a bounded value of $\lVert \pi\rVert_\infty$. The underlying reason for this extra assumption is the width $W$ of the transition window in \Cref{lem:cert}. Indeed, if we could take $W=1$ in the lemma, the extra assumption on the point probabilities of $\pi$ could be avoided.
This is not possible in general: the transition width in \Cref{lem:cert} is optimal up to a factor $\log n$ in the worst case. Let $P$ and $P'$ be the CDFs of $p$ and $p'$ from \Cref{prop:2Distros}, and assume $m<m_0$ and $\gamma\le1/4$, so that $P(m)\ge p_m\ge1-\gamma$ and $P'(m)\le1-p'_{m_0}\le\gamma$.
Suppose that a polynomial $q$ of degree at most $k$ satisfies properties (i)--(iv) of that lemma with $M=m$ and some integer width $W\ge1$ such that $m+W\le n-1$. The polynomial is feasible for the dual CDF problem [D$'$], so weak duality and properties (iii)--(iv) give
	\[
	P(m)\le \sum_{t=0}^{n-1}p'_t q(t)\le P'(m+W-1)+n^{-2},
	\]
	where we use that $p$ and $p'$ have the same first $k$ moments. If $m+W-1<m_0$, then $P'(m+W-1)\le1-p'_{m_0}\le\gamma$, whereas $P(m)\ge1-\gamma$. This would imply $1-\gamma\le\gamma+n^{-2}$, contradicting $\gamma\le1/4$ and $n\ge3$. Hence $W\ge m_0-m+1$. When $n/k$ is sufficiently large, \Cref{prop:2Distros} allows us to choose central peaks with $m_0-m=\Theta(n/k)$ while keeping $\gamma\le1/4$, proving $W=\Omega(n/k)$. %

\section{Proofs for Linear Sketches (\Cref{sec:linear-sketches})}
\label{app:linear-sketches}

We describe the linear-sketch model and give the proofs of the bounds stated in \Cref{sec:linear-sketches}. A $k$-dimensional linear sketch is specified by a matrix $A\in\mathbb R^{k\times n}$; on input distribution $p\in\Delta_n$, the sketch is the vector $Ap\in\mathbb R^k$. Equivalently, the sketch consists of $k$ inner products with fixed vectors. In this model we bound only the number of linear measurements, and not the bit complexity of the entries of $A$ or of the sketch values.

As in \Cref{sec:model-coarse-greedy}, the matrix $A$ defines the sketch $\sigma_A(p):=Ap$, and hence the uncertainty collection $\mathcal U_{\sigma_A}$, whose uncertainty sets are the sets $\{p\in\Delta_n: Ap=y\}$ for $y\in A\Delta_n$. An algorithm that only knows $Ap$ must use the same deterministic search strategy for every distribution in the uncertainty set containing $p$, and the best worst-case guarantee it can achieve is the approximability factor $\alpha(\mathcal U_{\sigma_A})$.

Linear sketches are also widely used in streaming algorithms, where storage is typically measured in bits~\cite{gilbert2002summarize,cormode2005improved}. The relationship between real-valued and finite-precision sketching is studied by Gribelyuk et al.~\cite{gribelyuk_lifting_2025}, who transfer dimension lower bounds for real-valued linear sketches to settings with polynomially bounded integer inputs and sketching matrices. In our setting, the bit lower bound applies to every encoding, linear or not. Thus, when each stored measurement uses $O(\log n)$ bits, it directly implies that $n^{\Omega(\eps)}$ measurements are necessary for an $O(1/\eps)$-approximation. The next subsection removes this restriction on encoding length and proves a lower bound directly for exact real-valued measurements.

\subsection{Lower Bound for Linear Sketches}

The next theorem shows that $\alpha(\mathcal U_{\sigma_A})=O(1/\varepsilon)$
requires a sketch of dimension $k\ge n^{\Omega(\varepsilon)}$. This lower bound is tight up to a constant in the exponent and polylogarithmic factors, as shown by
the dyadic Johnson--Lindenstrauss sketch of \Cref{subsec:dyadic-jl-sketch}.

\linearSketchLowerBound*
\begin{proof}
Let $\bar p$ be the uniform distribution on $[n]_0$, and let
$U:=\{p\in\Delta_n: Ap=A\bar p\}\in\mathcal U_{\sigma_A}$ be the uncertainty set
containing it. Fix any deterministic search strategy $S$ for $U$,
so that $\operatorname{cost}_p(S)=\sum_t p_t q(S,t)$.
We show that some $p^*\in U$ has
$\operatorname{cost}_{p^*}(S)/\opt(p^*)=\Omega(\log n/\log k)$.
By the entropy lower bound~\cite{mehlhornNearlyOptimalBinary1975},
\[
\sum_{t=0}^{n-1}\bar p_t\,q(S,t)
=\Omega(H(\bar p))=\Omega(\log n).
\]

Consider the linear program
\[
\begin{aligned}
\max \quad & \sum_{t=0}^{n-1} p_t\,q(S,t) \\
\text{s.t.}\quad
& \sum_{t=0}^{n-1} p_t=1,\\
& Ap=A\bar p,\\
& p_t\ge 0 \qquad \text{for all }t\in[n]_0.
\end{aligned}
\]
The uniform distribution $\bar p$ is feasible and has objective value
$\Omega(\log n)$.
Let $p^*$ be a vertex optimal solution. Its objective value is at least
that of $\bar p$. Since $Ap^*=A\bar p$, we have $p^*\in U$, and
\[
\operatorname{cost}_{p^*}(S)
=\sum_{t=0}^{n-1}p^*_t\,q(S,t)
\ge
\sum_{t=0}^{n-1}\bar p_t\,q(S,t)
=
\Omega(\log n).
\]

It remains to upper bound $\opt(p^*)$. The feasible polytope has at most $k+1$ independent equality constraints, so every vertex has at most $k+1$ positive coordinates. In particular, $p^*$ has support size at most $k+1$. A strategy knowing $p^*$ repeatedly queries the middle remaining support point, identifying every positive-probability target in $O(\log k)$ queries, since $k\ge2$. On reaching an interval containing no support point, it finishes by ordinary binary search.
Thus, $\opt(p^*)=O(\log k)$. Combining this with the lower bound
on $\operatorname{cost}_{p^*}(S)$ gives
\[
\frac{\operatorname{cost}_{p^*}(S)}{\opt(p^*)}
\ge
\Omega\left(\frac{\log n}{\log k }\right).
\]
Since $S$ was arbitrary, $\alpha(\mathcal U_{\sigma_A})=\Omega(\log n/\log k)$.
\end{proof}

\subsection{A Dyadic Johnson--Lindenstrauss Sketch for CDF Approximation}
\label{subsec:dyadic-jl-sketch}

We now give a simple linear sketch that approximately reconstructs the CDF of
every distribution. Together with \Cref{thm:CoarseGreedy}, this implies
that the lower bound of \Cref{thm:linearSketchlowerBound} is tight up to a constant in the exponent and polylogarithmic factors: a linear sketch $\sigma_V(p)=Vp$ of dimension $\widetilde O(n^\varepsilon)$
suffices for $\alpha(\mathcal U_{\sigma_V})=O(1/\varepsilon)$.
The construction uses the dyadic decomposition underlying
range-query and quantile sketches~\cite{gilbert2002summarize,cormode2005improved},
combined here with near-orthogonal vectors.
For simplicity, assume first that $n=2^m$; the general case follows
by padding to the next power of two.

For each level $\ell=0,\ldots,m$, let
$I^{(\ell)}_0,\ldots,I^{(\ell)}_{2^\ell-1}$ be the level-$\ell$ dyadic
partition of $[n]_0$, where
$I^{(\ell)}_a=\{a2^{m-\ell},a2^{m-\ell}+1,\ldots,(a+1)2^{m-\ell}-1\}$. Thus,
level $\ell$ partitions the path into $2^\ell$ consecutive intervals of equal
size. For a distribution $p$, let $z^{(\ell)}_a:=p(I^{(\ell)}_a)$ be the mass
of the dyadic interval $I^{(\ell)}_a$.

Instead of storing all dyadic masses exactly, we sketch them using projections.
For each level $\ell=0,\ldots,m$, consider vectors
$g^{(\ell)}_a\in\mathbb R^R$ that are near-orthogonal: given an error tolerance
$\eta>0$, they satisfy
\[
\left|
\langle g^{(\ell)}_a,g^{(\ell)}_b\rangle-\indic_{\{a=b\}}
\right|
\le \eta
\qquad\text{for all }a,b,\ell.
\]
The existence of such vectors is a direct consequence of the Johnson--Lindenstrauss lemma~\cite{johnson1984extensions}. Indeed, for each level $\ell$, apply a JL map to the set consisting of the origin and the canonical vectors $e_0,\ldots,e_{2^\ell-1}\in\mathbb R^{2^\ell}$, and let $g^{(\ell)}_a$ be the image of $e_a$. Preservation of distances in this set implies the above approximate inner-product identities. Since $2^\ell\le n$, it suffices to take $R=O(\eta^{-2}\log n)$. Fix such vectors independently of $p$. The error bounds below then hold simultaneously for every distribution $p\in\Delta_n$.

At level $\ell$, the sketch stores
\[
y^{(\ell)}:=\sum_{a=0}^{2^\ell-1} z^{(\ell)}_a g^{(\ell)}_a.
\]
Since $z^{(\ell)}_a$ is linear in $p$, the vector $y^{(\ell)}$ is also linear
in $p$. The full sketch is the block vector
$Vp:=(y^{(0)},y^{(1)},\ldots,y^{(m)})\in\mathbb R^{R(m+1)}$, for a suitable
matrix $V$. Thus the sketch has $R(m+1)=O(\eta^{-2}\log^2 n)$ coordinates.

The reconstruction is linear as well. To estimate the mass of
$I^{(\ell)}_a$, define
$\widetilde z^{(\ell)}_a:=\langle g^{(\ell)}_a,y^{(\ell)}\rangle$. Expanding,
\[
\widetilde z^{(\ell)}_a
=
\sum_{b=0}^{2^\ell-1}z^{(\ell)}_b
\langle g^{(\ell)}_a,g^{(\ell)}_b\rangle.
\]
Using the near-orthogonality of the vectors $g^{(\ell)}_a$, for every
distribution $p$ and every dyadic interval $I^{(\ell)}_a$, we have
$\left|\widetilde z^{(\ell)}_a-z^{(\ell)}_a\right|
\le \eta\sum_b z^{(\ell)}_b=\eta$, because the level-$\ell$ dyadic intervals
partition $[n]_0$.

It remains to reconstruct prefixes. Every prefix $\{0,\ldots,t\}$ has a
canonical decomposition into at most $m+1$ dyadic intervals, one per level.
Write $\{0,\ldots,t\}=J_1\cup\cdots\cup J_s$, with $s\le m+1$, where each
$J_r$ is some dyadic interval $I^{(\ell_r)}_{a_r}$. Since
$P(t)=\sum_{r=1}^s z^{(\ell_r)}_{a_r}$, define
\[
\widetilde P(t)
:=
\sum_{r=1}^s
\left\langle g^{(\ell_r)}_{a_r},y^{(\ell_r)}\right\rangle.
\]
This reconstruction is linear in the stored sketch $Vp$. Hence there is a
deterministic matrix $U$ such that $\widetilde P=U(Vp)$.

\begin{theorem}[Dyadic JL sketch for CDF approximation]
\label{thm:dyadicJL}
For every $\delta\in(0,1)$, there exist matrices
$U\in\mathbb R^{n\times k}$ and $V\in\mathbb R^{k\times n}$ with
$k=O(\log^4 n/\delta^2)$ such that every distribution $p\in\Delta_n$ satisfies
$\|U(Vp)-P\|_\infty\le \delta$, where $P$ is the CDF of $p$.
\end{theorem}

\begin{proof}
With the construction above, each dyadic mass estimate has error at most $\eta$. Therefore $|\widetilde P(t)-P(t)|\le (m+1)\eta$ for every $t\in[n]_0$. Choosing $\eta:=\delta/(m+1)$ gives a uniform CDF approximation with error at most $\delta$. The dimension is $k=(m+1)R=O(\eta^{-2}\log^2 n)=O(\log^4 n/\delta^2)$.

The construction and analysis above prove the result when $n$ is a power of
two. If $n$ is not a power of two, embed $[n]_0$ into $[N]_0$ with
$N=2^{\lceil\log_2 n\rceil}\le 2n$ by adding zero-probability dummy vertices after
$n-1$. Applying the power-of-two construction on $[N]_0$ and restricting the
reconstructed CDF to the original vertices gives the desired guarantee.
\end{proof}

Directly applying the previous sketch to our search algorithm \textsc{Coarse Greedy} in \Cref{thm:CoarseGreedy} yields the following corollary.
\linearSketchSearchUpperBound*

\begin{proof}
Apply \Cref{thm:dyadicJL} with $\delta=\tfrac12 n^{-\varepsilon/2}$. This gives matrices $U\in\mathbb R^{n\times k}$ and $V\in\mathbb R^{k\times n}$ with $k=O(n^\varepsilon\log^4 n)$ such that $\|U(Vp)-P\|_\infty\le\delta$ for every distribution $p\in\Delta_n$, where $P$ is its CDF.

Fix any nonempty uncertainty set $U_y:=\{p\in\Delta_n:Vp=y\}$ of $\mathcal U_{\sigma_V}$. Using only $y$, choose a nominal distribution $\pi\in U_y$ by solving the linear feasibility problem $V\pi=y$, $\pi\ge0$, and $\sum_t\pi_t=1$, and let $\Pi$ be its CDF. For every $p\in U_y$, the sketch guarantee applies to both $p$ and $\pi$, so
\[
d_K(p,\pi)=\|P-\Pi\|_\infty\le\|P-Uy\|_\infty+\|Uy-\Pi\|_\infty\le2\delta=n^{-\varepsilon/2}.
\]
Thus $U_y\subseteq B_K(\pi,n^{-\varepsilon/2})$. Running \textsc{Coarse Greedy} with parameter $\varepsilon/2$ and nominal distribution~$\pi$ gives an $O(1/\varepsilon)$-approximation for every $p\in U_y$ by \Cref{thm:CoarseGreedy}. Since $U_y$ was arbitrary, $\alpha(\mathcal U_{\sigma_V})=O(1/\varepsilon)$.
\end{proof}

\section{Randomized Extensions of the Lower Bounds}
\label{sec:randomized-lower-bounds}

For an uncertainty set $U$, a randomized search strategy is a
probability distribution $\mathcal{D}$ over correct deterministic search
strategies. The strategy $S\sim\mathcal{D}$ is sampled independently of
the target and without knowing which $p\in U$ is the true distribution.
The randomized approximability factor of an uncertainty collection is
\[
\alpha_{\mathrm{rand}}(\mathcal{U})
:=\sup_{U\in\mathcal{U}}\inf_{\mathcal{D}}\sup_{p\in U}
\frac{\EE_{S\sim\mathcal{D}}[\operatorname{cost}_p(S)]}{\opt(p)}.
\]
Thus $\alpha_{\mathrm{rand}}(\mathcal{U})\le\alpha(\mathcal{U})$;
the comparator $\opt(p)$ remains unchanged.

We prove the information-theoretic lower bound even when the sketch
is randomized. Let $R$ collect all random choices of the sketch and
search rule, including shared randomness, independently of the input
distribution and target. For fixed $R=r$, write $\sigma_r$ for the sketch
and $S_r(x)$ for the correct deterministic search strategy chosen on
message $x$. The search rule learns about $p$ only through
$\sigma_R(p)$.

\begin{theorem}[Randomized information-theoretic lower bound]
\label{thm:randomizedInformationLowerBound}
For every fixed $\eps\in(0,1/4)$, there exists $n_0(\eps)$ such that
the following holds for all $n\ge n_0(\eps)$.
Consider any randomized sketch and search rule $(\sigma_R,S_R)$ with
$\sigma_r:\Delta_n\to\{0,1\}^k$ for every $r$, where
$k\le\frac12 n^\eps\log n$.
Then there exists a distribution $p\in\Delta_n$ such that
\[
\EE_R\!\left[
\operatorname{cost}_p(S_R(\sigma_R(p)))
\right]
\ge \frac{\gamma_{\mathrm{rand}}}{\eps}\opt(p),
\]
where $\gamma_{\mathrm{rand}}>0$ is a universal constant.
\end{theorem}

\begin{proof}
Use the block family $p_Y$ from the proof of
\Cref{thm:informationLowerBound}, with $q=\lceil n^\eps\rceil$, and
choose $Y$ independently of $R$. Fix a seed $r$ and put
$X_r:=\sigma_r(p_Y)$. The same entropy calculation gives
$H(Y\mid X_r)\ge\frac18 q\log n$.
For each message $x$ of positive probability, the conditional-distribution
argument in that proof gives, for some constant $c>0$,
\[
\EE_{Y\mid X_r=x}[\operatorname{cost}_{p_Y}(S_r(x))]
\ge\frac{c}{q}H(Y\mid X_r=x).
\]
Here we average over all messages rather than selecting one. Consequently,
\[
\EE_Y[\operatorname{cost}_{p_Y}(S_r(X_r))]
\ge\frac{c}{q}H(Y\mid X_r)
\ge\frac{c}{8}\log n.
\]
Averaging also over $R$ yields
\[
\EE_Y\EE_R[\operatorname{cost}_{p_Y}(S_R(\sigma_R(p_Y)))]
\ge\frac{c}{8}\log n.
\]
Hence some fixed realization $y$, independent of $R$, has expected
search cost at least $(c/8)\log n$. As shown in the main proof,
$\opt(p_y)=O(\eps\log n)$, which gives the claimed ratio.
\end{proof}

In particular, for a deterministic sketch satisfying the hypotheses
of \Cref{thm:informationLowerBound},
$\alpha_{\mathrm{rand}}(\mathcal{U}_{\sigma})=\Omega(1/\eps)$.
The covering argument of \Cref{cor:Coveringlowerbound} applies unchanged
with $\alpha_{\mathrm{rand}}$ in place of $\alpha$.

\paragraph{The other lower bounds.}
The lower bounds for Kolmogorov balls, moments, and linear sketches
also hold with $\alpha_{\mathrm{rand}}$ in place of $\alpha$.
For \Cref{thm:kolmogorovBallLowerBound}, apply the randomized
covering bound to the quantile cover in \Cref{eq:kolmogorov-cover},
using the same parameter choice as in the main proof.
For \Cref{thm:momentSearchSeparation}, average the inequality
$(R+1)^{-1}\sum_{r=0}^R q(T,m_r)\ge\log_3(R+1)$ over the random tree
$T$; some fixed $r$ has this expected query cost, while
$\opt(p^{(r)})=O(1)$.
For \Cref{thm:linearSketchlowerBound}, replace the LP objective coefficients
$q(S,t)$ by $\bar q(t):=\EE_{S\sim\mathcal{D}}[q(S,t)]$.
The uniform distribution still has objective value $\Omega(\log n)$,
and an optimal vertex still has support size at most $k+1$.
Thus all three proofs retain their stated lower bounds.

\section{Sample Complexity for Search}
\label{subsec:sample-complexity}

Quantiles give an explicit bit-efficient representation of the distribution. If instead the algorithm has sample access to the distribution, then the empirical CDF gives another way to obtain the information needed by Coarse Greedy. We use the Dvoretzky--Kiefer--Wolfowitz inequality, in its sharp form due to Massart~\cite{dvoretzky1956asymptotic,massart1990tight}.

\begin{corollary}[Sample complexity for Coarse Greedy]
\label{cor:sampleComplexityCoarseGreedy}
Let $p\in\Delta_n$ and $\varepsilon,\delta\in(0,1)$. Suppose we draw
\[
N\ge \frac12 n^{\varepsilon}\ln\frac{2}{\delta}
\]
independent samples from $p$, and let $\widehat p_N$ be the empirical distribution. Run \textsc{Coarse Greedy} with nominal distribution $\widehat p_N$ and parameter $\varepsilon/2$. With probability at least $1-\delta$ over the samples, the resulting strategy $S$ satisfies
\[
\operatorname{cost}_p(S)\le O\left(\frac1\varepsilon\right)\operatorname{OPT}(p).
\]
In particular, for every fixed $\varepsilon>0$, this is a constant-factor approximation. The search target is drawn independently of the samples.
\end{corollary}

\begin{proof}
By the Dvoretzky--Kiefer--Wolfowitz inequality,
\[
\Pr\left(d_K(\widehat p_N,p)> n^{-\varepsilon/2}\right)
\le 2e^{-2N n^{-\varepsilon}}.
\]
The stated choice of $N$ makes the right-hand side at most $\delta$. Thus, with probability at least $1-\delta$, we have $d_K(\widehat p_N,p)\le n^{-\varepsilon/2}$. On this event, \Cref{thm:CoarseGreedy} with parameter $\varepsilon/2$ gives the claimed $O(1/\varepsilon)$-approximation.
\end{proof}

\section*{AI Disclosure} 
\addcontentsline{toc}{section}{AI Disclosure}

Generative AI tools (Claude Opus 5.5, Fable 5.1, and ChatGPT Astra) were used
to improve the presentation, including the generation of figures, as well as generate some of the proofs. The main questions and frameworks are
entirely human-generated. \Cref{lem:bump,lem:cert} had a human-generated previous version
based on Bernstein polynomials, which yielded the weaker bound
$d_K(\pi,p)=O\big(L\,\frac{n}{k^{1/3}}\big)$ in
\Cref{thm:momentRecoveryBoundedAtoms}. Similarly, \Cref{prop:2Distros} had a more
technical previous version, based on an explicit construction using Lagrange interpolation on Chebyshev nodes 
for~\Cref{eq:gammastar}, which required $(k+1)^4\le n$ and
gave $\gamma=O(k^3|m-m_0|/n)$. The current versions of these two results were 
proved by Fable 5.1. The proof of \Cref{thm:dyadicJL} was also AI-generated. All arguments have subsequently been verified, corrected, and rewritten by the authors. The authors take full responsibility for the content and correctness
of this work.

\section*{Acknowledgements}
\addcontentsline{toc}{section}{Acknowledgements \& References}
Part of this work was done during a sabbatical visit at Columbia University by the first and third authors. 
Nicole Megow was supported by DFG grants no.\ 547924951 and 517912373 and the Data Science Center of the University of Bremen (DSC@UB) funded by the State of Bremen.
Cliff Stein was supported in part  by NSF grant CCF-2218677, ONR grant ONR-13533312,
and the Wai T. Chang Chair in Industrial Engineering and Operations Research.
José Verschae was partially supported by ANID FONDECYT Regular grant no.\ 1260640 and by Centro de Modelamiento Matem\'atico (CMM), BASAL fund FB210005, ANID-Chile.

\bibliographystyle{alphaurl}
\bibliography{literature}

\end{document}